\documentclass[12pt]{article}%
\usepackage{amsmath}%
\usepackage{amsfonts}%
\usepackage{amssymb}%
\usepackage{graphicx}
\newtheorem{acknowledgement}{Acknowledgement}

\begin{document}

\hfill FIAN-TD-2012-39

\vspace{3cm}

\begin{center}

{\bf \Large Generalized oscillator representations

\medskip
for Calogero Hamiltonians}

\vspace{2cm}

I.V. Tyutin\footnote{e-mail:tyutin@lpi.ru}
and B.L. Voronov\footnote{e-mail:voronov@lpi.ru}

\vspace{1cm}

{\it Department of Theoretical Physics, P.N. Lebedev Physical Institute, 

\medskip
Leninsky prospect 53, Moscow, 119991, Russia}

\vspace{2cm}

{\bf Abstract}

\end{center}

This paper is a natural continuation of the previous paper J.Phys. A:
Math.Theor. 44 (2011) 425204, arXiv:0907.1736 [quant-ph]
where oscillator representations for nonnegative Calogero Hamiltonians with
coupling constant $\alpha\geq-1/4$ were constructed. Here, we present
generalized oscillator representations for all Calogero Hamiltonians with
$\alpha\geq-1/4$.These representations are generally highly nonunique, but
there exists an optimum representation for each Hamiltonian.

\newpage
\section{Introduction}

This paper is a natural continuation of the previous papers \cite{GitTyV10}
and \cite{GitTyV11}. In \cite{GitTyV10}, see also sec. 7.2 in \cite{GitTyV12}
to which we mainly refer in what follows, all one-particle Calogero
Hamiltonians $\hat{H}_{\mathfrak{e}}$\thinspace associated with self-adjoint
(s.a. in what follows) Calogero differential operation\footnote{We recall that
by definition, a differential operator $\hat{f}$ is called associated with a
differential operation $\check{f}$ if the operator $\hat{f} $ acts on its
domain $D_{f\text{ }}$ by $\check{f}$: $\hat{f}\psi(x)=$ $\check{f}$
$\psi(x),\forall\psi\in D_{f\text{ }}$.}%

\begin{equation}
\check{H}=-d_{x}^{2}-\alpha/x^{2},x\in\mathbb{R}_{+},\text{ }\alpha
\in\mathbb{R},\label{2.1}%
\end{equation}
$\alpha$ is a dimensionless coupling constant, were constructed as \ s.a.
operators in $L^{2}(\mathbb{R}_{+})$ and their spectra and (generalized)
eigenfunctions were evaluated including inversion formulas. In \cite{GitTyV11}%
, the so-called oscillator representations for nonnegative Calogero
Hamiltonians $\hat{H}_{\mathfrak{e}}$, $\hat{H}_{\mathfrak{e}}\geq0 $, with
$\alpha\geq-1/4$ were constructed. An oscillator representation is a
representation of the form
\begin{equation}
\hat{H}_{\mathfrak{e}}=\hat{c}^{+}\hat{c},\label{1.2}%
\end{equation}
where $\hat{c}$ and $\hat{c}^{+}$is a pair of closed mutually adjoint
first-order differential operators,$\ \hat{c}^{+}=(\hat{c})^{+},\,\hat{c}$
$=\overline{c}$ $=(\hat{c}^{+})^{+}$. Such a representation makes
evident\textrm{\ }that the Hamiltonian $\hat{H}_{\mathfrak{e}}$ is
nonnegative. Here, the results in \cite{GitTyV11} are generalized to all
Calogero Hamiltonians with $\alpha\geq-1/4$, generally not nonnegative, in the
form of generalized oscillator representations.

The initial basic ideas of the generalization are as follows.

As is known, all Calogero Hamiltonians with $\alpha\geq-1/4$ are bounded from
below \cite{GitTyV12}. From the general standpoint, this is a consequence of
that the initial symmetric operator $\hat{H}$ associated with $\check{H}$ with
$\alpha\geq-1/4$ and defined on the subspace $\mathcal{D}(\mathbb{R}_{+})$ of
smooth compactly supported functions is nonnegative and therefore, all its
s.a. extensions $\hat{H}_{\mathfrak{e}}$, which are just the Calogero
Hamiltonians, are bounded from below \cite{AkhGl81, Naima69}. This implies
that for each Calogero Hamiltonian $\hat{H}_{\mathfrak{e}}$ with $\alpha
\geq-1/4$, there exists a nonnegative constant $u$ such that the operator
$\hat{H}_{\mathfrak{e}}+u^{2}\hat{I}$, where $\hat{I}$ is the identity
operator, is nonnegative and may allow an oscillator representation of form
(\ref{2.1}). The operator $\hat{H}_{\mathfrak{e}}+u^{2}\hat{I}$ is evidently
associated with the differential operation $\check{H}+u^{2}$. The parameter
$u$ is of dimension of inverse length, and it is convenient to represent it as
$u=sk_{0}$, where $s\geq0$ is a dimensionless parameter and $k_{0}$ $>0$ is a
fixed parameter of dimension of inverse length. By a generalized oscillator
representation for Calogero Hamiltonian $\hat{H}_{\mathfrak{e}}$, we mean a
representation of the form%
\begin{equation}
\hat{H}_{\mathfrak{e}}=\hat{c}^{+}\hat{c}-(sk_{0})^{2}\hat{I},\,s\geq
0,\label{1.3}%
\end{equation}
where $\hat{c}$ and $\hat{c}^{+}$are of the previous meaning; $s=0$ yields an
oscillator representation. Such representations is one more aspect of the
general Calogero problem. They can be useful for many reasons, including the
spectral analysis of the Hamiltonians. In particular, representation
(\ref{1.3}) for a Hamiltonian $\hat{H}_{\mathfrak{e}}$ makes
evident\textrm{\ }that\textrm{\ }$\hat{H}_{\mathfrak{e}}$ is bounded from
below, \thinspace and its spectrum is bounded from below by $-(sk_{0})^{2}$,
which is the lower boundary of the spectrum if the kernel of the
operator\textrm{\ }$\hat{c}$ is nontrivial, \textrm{ker}$\hat{c}\neq\{0\}$;
then \textrm{ker}$\hat{c}$ is the groundspace (ground state) of the
Hamiltonian and $E_{0}=-(sk_{0})^{2} $ is its ground-state energy.

A starting point for constructing oscillator representations (\ref{1.2}) was
the oscillator representation for the respective differential operation
$\check{H}$ that is a representation of the form
\begin{align*}
\check{H}  &  =\check{b}\check{a},\\\check{b}
&  =\check{a}^{\ast}=(\check{b}^{\ast})^{\ast},\,\check{a}=\ \check%
{b}^{\ast}=(\check{a}^{\ast})^{\ast},
\end{align*}
where $\check{a}$ is a first-order differential operation and $\check{a}%
^{\ast}$is its adjoint by Lagrange, see \cite{AkhGl81}. Accordingly, a
starting point for constructing generalized oscillator representations
(\ref{1.3}) for Calogero Hamiltonians with the coupling constant $\alpha
\geq-1/4 $ should be a generalized oscillator representation%
\begin{equation}
\check{H}=\check{b}\check{a}-(sk_{0})^{2},\,s\geq0,\label{1.5}%
\end{equation}
for the respective $\check{H}$ with certain $\check{a}$\ and\textrm{\ }%
$\check{b}$\textrm{\ }of the previous meaning.

Let differential operation $\check{H}$ (\ref{2.1}) allows generalized
oscillator representation (\ref{1.5}). If we introduce the pair of initial
differential operators $\hat{a}$ and $\hat{b}$ in $L^{2}(R_{+})$%
\textrm{\thinspace}defined on $\mathcal{D}(\mathbb{R}_{+})$ and associated
with the pair of the respective differential operations $\check{a}$ and
$\check{b}$, then the initial symmetric operator $\hat{H}$ is evidently
represented as%

\[
\hat{H}=\hat{b}\,\hat{a}-(sk_{0})^{2}\hat{I}.
\]
Let $\hat{c}$ and $\hat{c}^{+}$ be a pair of closed mutually adjoint operators
that are closed extentions of the respective initial operators $\hat{a}$ and
$\hat{b}$, $\hat{a}\subset\hat{c}$, $\hat{b}\subset\hat{c}^{+} $. Then the
operator
\begin{equation}
\hat{H}_{ext}=\hat{c}^{+}\hat{c}-(sk_{0})^{2}\hat{I}\label{1.7}%
\end{equation}
is an evident extension of $\hat{H}$, $\hat{H}\subset\hat{H}_{ext}$. By the
von Neumann theorem \cite{Neuma50}, for a proof, see also \cite{AkhGl81}, the
operator $\hat{N}=\hat{c}^{+}\hat{c}$, where $\hat{c}$ is closed, $\hat{c}$
$=\overline{\hat{c}}$, is s.a. and nonnegative, $\hat{N}\ =\hat{N}^{+}\geq0$;
in addition, if $\ker\hat{c}\neq\{0\}$, then $\hat{c}$ is an eigenspace \ with
the minimum eigenvalue $0$ which is the lower boundary of the spectrum of
$\hat{N}$. Therefore, operator $\hat{H}_{ext}$ (\ref{1.7}) is s.a. , which
means that $\hat{H}_{ext}$ is a certain Calogero Hamiltonian
$\hat{H}_{\mathfrak{e}}$ represented in the generalized oscillator form
(\ref{1.3}) providing its above-mentioned properties.

We note that generalized oscillator representation (\ref{1.3}) for a Calogero
Hamiltonian is equivalent to the representation%

\[
\hat{H}_{\mathfrak{e}}=\hat{d}\,\hat{d}^{+}-(sk_{0})^{2}\hat{I},
\]
where $\hat{d}$ and$\,\hat{d}^{+}$is a pair of closed mutually adjoint
operators that are extensions of the respective initial operators $\hat{b}$
and $\hat{a}$: it is sufficient to make the identifications $\,\hat{c}=\hat
{d}^{+},\,\hat{c}^{+}=\hat{d}$. Constructing a pair $\hat{c}\supset$
$\hat{a}$, $\hat{c}^{+}\supset$ $\hat{b}$ or a pair $\hat{d}\supset$ $\hat{b}%
$, $\,\hat{d}^{+}\supset$ $\hat{a}$ is a matter of convenience: we can start
with extending $\hat{a}$ to its closure or with extending $\hat{b}$ to its closure.

Varying the parameter $s$ in (\ref{1.5}) and involving all possible
mutually-adjoint extensions of the initial $\hat{a}$ and $\hat{b}$ with given
$s$, we can hope to construct generalized oscillator representations
(\ref{1.3}) for all Calogero Hamiltonians with $\alpha\geq-1/4\,$. We show
below that these expectations are realized. An identification of Hamiltonians
$\hat{H}_{\mathfrak{e}}$ (\ref{1.3}) with the known Calogero Hamiltonians in
\cite{GitTyV12} is straightworward for $\alpha\geq3/4$ because the Hamiltonian
with given $\alpha\geq3/4$ is unique, while for $-1/4\leq\alpha<3/4$, an
identification goes through evaluating the asymptotic behavior of functions
belonging to the domain of $\hat{H}_{\mathfrak{e}}$ (\ref{1.3})\ at the origin
and comparing it with the asymptotic s.a. boundary conditions specifying
different Calogero Hamiltonians with\ given $\alpha\in\lbrack-1/4,3/4)$
\cite{GitTyV12}.

We say in advance that generalized oscillator representation (\ref{1.3}) for a
given Calogero Hamiltonian is generally highly nonunique; in fact, there
exists a one-, or even two-, parameter family of generalized oscillator
representations for each Hamiltonian, among which there exists an optimum representation.

As to generalized oscillator representations for Calogero Hamiltonians with
the coupling constants $\alpha<-1/4$, there are no such representations and
can not be because these Hamiltonians are not bounded from below
\cite{GitTyV12}.\textrm{\ }

\section{General}

We begin with looking into the possibility of generalized oscillator
representation (\ref{1.5}) for Calogero differential operation $\check{H}$
(\ref{2.1}), i.e., representing\textrm{\ }$\check{H}$ as a product of two
finite-order differential operations\textrm{\ \ }mutually adjoint by Lagrange
minus a nonnegative constant:%

\begin{align}
\check{H} &  =\check{b}(s)\,\check{a}(s)-(sk_{0})^{2},\,s\geq0,\nonumber\\
\check{b} &  =\check{a}^{\ast}=(\check{b}^{\ast})^{\ast},\,\check{a}%
=\ \check{b}^{\ast}=(\check{a}^{\ast})^{\ast},\label{2.3}%
\end{align}
the fixed parameter $k_{0}$ is of dimension of inverse length. In fact, we
deal with a family $\{\check{a}(s),$ $\check{b}(s)\}$ of mutually adjoint by
Lagrange differential operations. We often omit the argument $s$ in the
symbols $\check{a}$ and $\check{b}$ for brevity and write it\ when needed. The
representation (\ref{2.3}) with $s=0$ is an oscillator representation for
$\check{H}$, it was considered in \cite{GitTyV11}. It is desirable that
$\check{a}(s)$ and $\check{b}(s)$ be continuous in $s$, in particular, it is
desirable to get the known oscillator representations for $\check{H}$ from
generalized oscillator representations (\ref{2.3}) for $\check{H}$\ in the
limit $s\rightarrow0$.

It is easy to see that $\check{a}$ and $\check{b}$ have to be first-order
differential operations,%
\begin{equation}
\check{a}=e^{i\theta(x)}[d_{x}-h(x)],\;\check{b}=-[d_{x}+\overline
{h(x)}]e^{-i\theta(x)}.\label{2.5}%
\end{equation}
The substitution of (\ref{2.5}) into (\ref{2.3}) results in the following
necessary and sufficient conditions on the function $h(x)$ for representation
(\ref{2.3}) to hold:%

\[
\operatorname{Im}h(x)=0,\;h^{^{\prime}}(x)+h^{2}(x)=\alpha x^{-2}+s^{2}%
k_{0}^{2},
\]
which is the Ricatti equation. We additionally require that the functions
$\theta(x)$ and $h(x)$ be nonsingular in $(0,\infty)$. The arbitrary phase
factors $e^{\pm i\theta(x)}$ in (\ref{2.5}) are irrelevant because they
trivially cancel in the product $\check{b}\,\check{a}$; their fixing is a
matter of convenience. We set $\theta(x)=0$, so that in what follows,
$\check{a}$ and $\check{b}$ in representation (\ref{2.3}) are given by%
\begin{align}
\check{a}  &  =d_{x}-h(x),\,\check{b}=-d_{x}-h(x),\label{2.7a}\\
\operatorname{Im}h(x)  &  =0,\;h^{^{\prime}}(x)+h^{2}(x)=\alpha x^{-2}%
+s^{2}k_{0}^{2}.\label{2.7b}%
\end{align}
We recall that in fact, we have a family $\{\check{a}(s)=d_{x}-h(s;x),$
$\check{b}(s)=-d_{x}-h(s;x)\}$\ of mutually adjoint first-order differential
operations and the corresponding family $\{h(s;x)\}$ of functions $h$.

The representation (\ref{2.3}), (\ref{2.7a}), (\ref{2.7b}) with given $s$, if
it exists, is generally nonunique: if nonlinear equation (\ref{2.7b}) with
given $s$ has a family of different admissible solutions $h$, there is a
family of the respective different pairs $\check{a},\check{b}$ (\ref{2.7a})
providing the desired representation, so that apart from $s$, the symbols
$\check{a}$ and $\check{b}$ can contain a certain additional argument, let it
be $\mu$, parametrizing the family of admissible $h$'s with given $s$, and we
actually have a two-parameter family $\{h(\mu,s;x)\}$ and the respective
family $\{\check{a}(\mu,s),\,\check{b}(\mu,s)\}$. Where possible, this
parametrization must be such that to provide a smooth transition to the limit
$s\rightarrow0$ which reproduces the known oscillator representations for
$\check{H}$. We write the arguments $s$ and $\mu$ when needed and omit them
for brevity if this does not lead to misunderstanding.

It is easy to prove that differential operation $\check{H}$ (\ref{2.1}) allows
generalized oscillator representation (\ref{2.3}), (\ref{2.7a}), (\ref{2.7b}),
iff the homogeneous differential equation
\begin{equation}
-\phi^{\prime\prime}(x)+\frac{\alpha}{x^{2}}\phi(x)+(sk_{0})^{2}%
\phi(x)=0\label{2.8}%
\end{equation}
or the eigenvalue problem
\[
\check{H}\phi=-\phi^{\prime\prime}(x)+\frac{\alpha}{x^{2}}\phi(x)=-(sk_{0}%
)^{2}\phi(x),
\]
(which can be considered a stationary Schroedinger equation \ with ''energy''
$-(sk_{0})^{2}$) has a real-valued positive solution (with no zeroes in
$(0,\infty)$),%
\[
\operatorname{Im}\phi(x)=0,\;\phi(x)>0,\;x>0,
\]
and in this case,
\[
h(x)=\phi^{\prime}(x)/\phi(x)=-\phi(x)\left(  \frac{1}{\phi(x)}\right)
^{\prime},
\]
so that $\check{a}$ and $\check{b}$ allow the representations%
\begin{equation}
\check{a}=\phi(x)d_{x}\frac{1}{\phi(x)},\ \check{b}=-\frac{1}{\phi(x)}%
d_{x}\phi(x).\label{2.12}%
\end{equation}
It is evident that the function $\phi$ is defined up to a positive constant factor.

We also note that by $\phi(x)$ is actually meant a family $\{\phi(s;x)\}$, we
write the argument $s$ when needed.

Necessity. Let $\check{H}$ allow representation (\ref{2.3}), (\ref{2.7a}),
(\ref{2.7b}) with a function $h(x)$ absolutely continuous in $(0,\infty)$. We
introduce the real-valued positive function $\phi(x)$ defined up to a positive
constant factor by%
\[
\phi(x)=\exp\int^{x}d\xi\,h(\xi),
\]
so that the function $h(x)$ can be represented as
\[
h(x)=\phi^{\prime}(x)/\phi(x)=-\phi(x)\left(  \frac{1}{\phi(x)}\right)
^{\prime}.
\]
It is \ easy to verify that equation (\ref{2.7b}) for $h(x)$ implies eq.
(\ref{2.8}) for $\phi(x)$.

Sufficiency. Let $\phi(x)$ be a real-valued positive solution of eq.
(\ref{2.8})\ . It is easy to verify that the function $h(x)=$ $\phi^{\prime
}(x)/\phi(x)$ is absolutely continuous in $(0,\infty)$ and satisfies eq.
(\ref{2.7b}) thus providing representation (\ref{2.3}), (\ref{2.7a}),
(\ref{2.7b}) for $\check{H}$.

We thus obtain that an existence and a possible nonuniqueness of
representation (\ref{2.3}) is formulated in terms of eq. (\ref{2.8}) as
follows. If eq. (\ref{2.8}) has no real-valued positive solution, there exists
no representation (\ref{2.3}). If eq. (\ref{2.8}) has a unique, up to a
positive constant factor, real-valued positive solution $\phi$, there exists
unique representation (\ref{2.3}), (\ref{2.12}). If eq. (\ref{2.8}) has two
linearly independent real-valued positive solutions $\phi_{1}$ and $\phi_{2}$,
there exists a one-parameter family $\{\check{a},\check{b}\}$ of different
admissible pairs $\check{a},\check{b}$ providing the desired representation.
This family is in one-to-one correspondence given by (\ref{2.12}) with a
family $\{A\phi_{1}+B\phi_{2},\,A,B:A\phi_{1}(x)+B\phi_{2}(x)>0,\,x>0\}$ of
pairwise linearly independent real-valued positive solutions of eq.(\ref{2.8})
defined modulo a positive constant factor: the constant coefficients $A$ and
$B$ that differ by a positive constant factor yield the same pair $\check
{a},\check{b}$. The range of admissible coefficients $A$ and $B$ has to be
established. When possible, a single parameter, let it be $\mu$, parametrizing
these coefficients, $A=A(\mu)$ and $B=B(\mu)$, must be introduced in such a
way as to provide a proper oscillator representations for $\check{H}$ in the
limit $s\rightarrow0$.

The general solution of eq. (\ref{2.8}) with $s>0$ is given by%

\begin{align}
&  \phi(s;x)=A\sqrt{x}I_{\varkappa}(sk_{0}x)+B\sqrt{x}K_{\varkappa}%
(sk_{0}x),\nonumber\\
&  \varkappa=\left\{
\begin{array}
[c]{l}%
\sqrt{\alpha+1/4}\geq0,\;\alpha\geq-1/4\\
i\sigma,\;\sigma=\sqrt{|\alpha|-1/4}>0,\;\alpha<-1/4
\end{array}
\right.  ,\label{2.15}%
\end{align}
or by%
\begin{equation}
\phi(s;x)=A\sqrt{x}I_{\varkappa}(sk_{0}x)+B\sqrt{x}I_{-\varkappa}%
(sk_{0}x),\label{2.16}%
\end{equation}
if $\varkappa\notin Z_{+}$.

Here, $I_{\pm\varkappa}$ are the modified first-order Bessel functions and
$K_{\varkappa}$ is the McDonald function (another name for these functions is
the Bessel functions of imaginary argument ), $A$ and $B$ are arbitrary
complex constants. Whether the right-hand sides in (\ref{2.15}) or in
(\ref{2.16}) can be real-valued and positive under an appropriate choice of
coefficients $A$ and $B$ crucially depends on a value of the coupling constant
$\alpha$. Two regions of the coupling constant\thinspace differ drastically,
$\alpha<-1/4$ and $\alpha\geq-1/4$, which we consider separately. In the
second region, the point $\alpha=-1/4$ is naturally distinguished.

We first consider the region $\alpha<-1/4$ where a situation with generalized
oscillator representations is most simple.

\section{Region$\,\alpha<-1/4$ ($\varkappa=i\sigma$)}

In this region of the coupling constant, we use form (\ref{2.16}) with
$\varkappa=i\sigma$ of\ the general solution of eq. (\ref{2.8}) with $s>0$,%
\begin{equation}
\phi(s;x)=A\sqrt{x}I_{i\sigma}(sk_{0}x)+B\sqrt{x}I_{-i\sigma}(sk_{0}%
x),\;\sigma=\sqrt{|\alpha|-1/4}>0.\label{3.1}%
\end{equation}
Because the functions $I_{\pm i\sigma}$\ of real argument are linearly
independent and complex conjugate, $I_{-i\sigma}(sk_{0}x)=\overline
{I_{i\sigma}(sk_{0}x)}$,\ the condition $\operatorname{Im}\phi(sk_{0};x)=0$
requires that $B=\overline{A}$, and we obtain that the general real-valued
solution of eq. (\ref{2.8}), $s>0$, is given by%
\[
\phi(s;x)=2\operatorname{Re}[A\sqrt{x}I_{i\sigma}(sk_{0}x)].
\]

The asymptotic behavior of such a function as $x\rightarrow0$ is of the form%
\begin{align*}
&  \phi(s;x)=2\operatorname{Re}\left\{  \rho e^{i\varphi}\sqrt{x}%
(sk_{0}x)^{i\sigma}\left[  1+O(x^{2})\right]  \right\}  =\\
&  \,=2\sqrt{x}\rho\left[  1+O(x^{2})\right]  \left[  \cos(\sigma\ln
(sk_{0}x)+\varphi+O(x^{2})\right]  ,~\rho e^{i\varphi}\;=A\frac{(1/2)^{i\sigma
}}{\Gamma(1+i\sigma)},
\end{align*}
which demonstrates that any real-valued solution of eq. (\ref{2.8}) with
$\alpha<-1/4$ and $s>0$ has an infinite number of zeros in any neighbohood of
the origin. This means that\textrm{\ }the Calogero differential operation
$\check{H}$ (\ref{2.1}) with $\alpha<-1/4$ does not allow generalized
oscillator representation (\ref{2.3}) with $s>0$.

We recall that the same holds for an oscillator representation for $\check{H}
$, which corresponds to the case $s=0$, see \cite{GitTyV11}. \ We note that to
have the general solution of eq. $\check{H}\phi=0$ from the general solution
(\ref{3.1}) of eq. (\ref{2.8}) with $s\neq0$ as a proper limit $s\rightarrow
0$, it is sufficient to make the substitutions $A\rightarrow As^{-i\sigma}$,
$B\rightarrow Bs^{i\sigma}$.

\section{Region $\alpha>-1/4$ ($\varkappa>0$)}

\subsection{Generalized oscillator representations for $\check{H}$,
differential operations $\check{a}$ and $\check{b}$}

In this region of the coupling constant, we use form (\ref{2.15}) of the
general solution of eq. (\ref{2.8}) with $s>0$. To include the point $s=0$ in
the range of admissible $s,$ $s\geq0$, smoothly, it is appropriate to make the
substitutions \textrm{\ }
\[
A\rightarrow A\,\Gamma(1+\varkappa)\left(  \frac{s}{2}\right)  ^{-\varkappa
}\sqrt{k_{0}},\;B\rightarrow B\frac{2}{\Gamma(\varkappa)}\left(  \frac{s}%
{2}\right)  ^{\varkappa}\sqrt{k_{0}}.
\]
Under these substitutions, the general solution of eq. $\check{H}\phi=0$ is
properly obtained from the general solution (\ref{2.15}) of eq. (\ref{2.8})
with $s\neq0$ in the limit $s\rightarrow0$, which allows to get the known
oscillator representations for $\check{H}$\ (\ref{2.1}), see \cite{GitTyV11},
from generalized oscillator representations (\ref{2.3}), (\ref{2.12}) with
$s>0$ in the limit $s\rightarrow0$.

The general solution of eq. (\ref{2.8}) with $\alpha>-1/4$ and $s\geq0$, is
then given by
\begin{align}
&  \phi(s;x)=A\,\Gamma(1+\varkappa)\left(  \frac{s}{2}\right)  ^{-\varkappa
}\sqrt{k_{0}x}I_{\varkappa}(sk_{0}x)+B\frac{2}{\Gamma(\varkappa)}\left(
\frac{s}{2}\right)  ^{\varkappa}\sqrt{k_{0}x}K_{\varkappa}(sk_{0}%
x),\nonumber\\
&  \varkappa=\sqrt{\alpha+1/4}>0.\label{4.1.2}%
\end{align}
Because $\sqrt{k_{0}x}I_{\varkappa}(sk_{0}x)$ and $\sqrt{k_{0}x}K_{\varkappa
}(sk_{0}x)$ are real-valued and linearly independent solutions, a condition
$\operatorname{Im}\phi(s;x)=0$ requires that $\operatorname{Im}%
A=\operatorname{Im}B=0$. The function $I_{\varkappa}(sk_{0}x)$ is positive in
$(0,\infty)$ and monotonically increases from zero at $x=0$ to infinity as
$x\rightarrow\infty$ (see \cite{GraRy94} 8.445, 8.451.5), while the function
$K_{\varkappa}(sk_{0}x)$ is positive in $(0,\infty)$ and monotonically
decreases from infinity at $x=0$ to zero as $x\rightarrow\infty$ (see
\cite{GraRy94}, 8.432.1, 8.486.16, 8.486.11, 8.451.6, 8.485, 8.446).

It follows that $\phi(s;x)$ (\ref{4.1.2}) is real valued and positive in
$(0,\infty)$ iff $A\geq0$, $B\geq0$, $A+B>0$. As noted above, a common
constant positive factor in $\phi(s;x)$ is irrelevant because it does not
enter generalized oscillator representation (\ref{2.3}), (\ref{2.12}) for
$\check{H}$.

To make this evident, we can set $A=\sin\mu$, $B=\cos\mu,$\ $\mu\in
\lbrack0,\pi/2]$, in (\ref{4.1.2})\ without loss of generality.

Thus, for each $\alpha>-1/4$, $(\varkappa>0)$, we have a one-parameter family
$\{\phi(\mu,s;x)\}$ of pairwise linearly independent real-valued positive
solutions $\phi(\mu,s;x)$ of eq. (\ref{2.8}) with given $s\geq0$ and the
corresponding two-parameter family $\{\check{a}(\mu,s),\mathrm{\ }\check
{b}(\mu,s)\}$ of different pairs of mutually adjoint first-order differential
operations $\check{a}(\mu,s)$ and$\mathrm{\ }\check{b}(\mu,s)$ given by
(\ref{2.12}) with $\phi=\phi(\mu,s;x)$,%

\begin{align}
&  \check{a}(\mu,s)=\check{b}^{\ast}(\mu,s)=d_{x}-h(\mu,s;x)=\phi
(\mu,s;x)d_{x}\frac{1}{\phi(\mu,s;x)},\ \nonumber\\
&  \check{b}(\mu,s)=\check{a}^{\ast}(\mu,s)=-d_{x}-h(\mu,s;x)=-\frac{1}%
{\phi(\mu,s;x)}d_{x}\phi(\mu,s;x),\nonumber\\
&  h(\mu,s;x)=\frac{\phi^{\prime}(\mu,s;x)}{\phi(\mu,s;x)},\nonumber\\
&  \phi(\mu,s;x)=\sin\mu\,\Gamma(1+\varkappa)\left(  \frac{s}{2}\right)
^{-\varkappa}\sqrt{k_{0}x}I_{\varkappa}(sk_{0}x)+\cos\mu\frac{2}%
{\Gamma(\varkappa)}\left(  \frac{s}{2}\right)  ^{\varkappa}\sqrt{k_{0}%
x}K_{\varkappa}(sk_{0}x),\nonumber\\
&  \mu\in\lbrack0,\frac{\pi}{2}],\,\,s\in\lbrack0,\infty),\,\varkappa
>0,\label{4.1.4}%
\end{align}
providing a two-parameter family of different generalized oscillator
representations (\ref{2.3}) for $\check{H}$ (\ref{2.1}) with given
$\alpha>-1/4$,%
\begin{equation}
\check{H}=\check{b}(\mu,s)\check{a}(\mu,s)-(sk_{0})^{2}.\label{4.1.5}%
\end{equation}

As a rule, we indicate the ranges of parameters $\mu$ and $s$ in formulas to
follow only if they differ from the whole ranges, these are $[0,\pi/2]$ for
$\mu$ and $[0,\infty)$ for $s$, the range of $\varkappa$ is clear
from\textrm{\ }the title of section, subsection or subsubsection.

\textrm{\ }As to the main resulting formulas, we indicate the ranges of all
the parameters including\textrm{\ }$\varkappa$.

For the asymptotic behavior of the functions $\phi(\mu,s;x)$ and the functions
$1/\phi(\mu,s;x)$ at the origin, which we need below, we have%

\begin{align}
\phi(\mu,s;x)  &  =\left\{
\begin{array}
[c]{l}%
\left\{
\begin{array}
[c]{l}%
\cos\mu\,(k_{0}x)^{1/2-\varkappa}\left[  1+O(x^{2})\right]  ,\,\mu\in
\lbrack0,\pi/2)\\
(k_{0}x)^{1/2+\varkappa}\left[  1+O(x^{2})\right]  =O(x^{1/2+\varkappa}%
),\,\mu=\pi/2,
\end{array}
\right.  ,\;\varkappa>1,\\
\left\{
\begin{array}
[c]{l}%
\cos\mu\,(k_{0}x)^{-1/2}+O(x^{3/2}\ln\frac{1}{x}),\;\mu\in\lbrack0,\pi/2)\\
(k_{0}x)^{3/2}\left[  1+O(x^{2})\right]  =O(x^{3/2}),\,\,\mu=\pi/2,
\end{array}
\right.  ,\;\varkappa=1,\\
\tilde{A}(\mu,s)(k_{0}x)^{1/2+\varkappa}\left[  1+O(x^{2})\right]  +\cos
\mu\,(k_{0}x)^{1/2-\varkappa}\left[  1+O(x^{2})\right]  ,\;\\
\,\mu\in\lbrack0,\pi/2],\,0<\varkappa<1,
\end{array}
\right.  ,\;x\rightarrow0,\nonumber\\
\tilde{A}(\mu,s)  &  =\sin\mu-\frac{\Gamma(1-\varkappa)}{\Gamma(1+\varkappa
)}\left(  \frac{s}{2}\right)  ^{2\varkappa}\cos\mu\,,\,\label{4.1.6}%
\end{align}

and%

\begin{equation}
\frac{1}{\phi(\mu,s;x)}=\left\{
\begin{array}
[c]{l}%
\left\{
\begin{array}
[c]{l}%
\frac{1}{\cos\mu}(k_{0}x)^{\varkappa-1/2}\left[  1+O(x^{2})\right]  ,\,\mu
\in\lbrack0,\pi/2)\\
(k_{0}x)^{-1/2-\varkappa}\left[  1+O(x^{2})\right]  ,\;\,\mu=\pi/2
\end{array}
\right.  ,\;\varkappa>1\\
\left\{
\begin{array}
[c]{l}%
\frac{1}{\cos\mu}(k_{0}x)^{1/2}+O(x^{5/2}\ln\frac{1}{x}),\,\,\mu\in
\lbrack0,\pi/2)\,\\
(k_{0}x)^{-3/2}\left[  1+O(x^{2})\right]  ,\;\,\mu=\pi/2,\,
\end{array}
\right.  ,\;\varkappa=1\\
\left\{
\begin{array}
[c]{l}%
\frac{(k_{0}x)^{\varkappa-1/2}}{\cos\mu\,+\tilde{A}(\mu,s)(k_{0}%
x)^{2\varkappa}}\left[  1+O(x^{2})\right]  ,\;\mu\in\lbrack0,\pi/2),\\
(k_{0}x)^{-1/2-\varkappa}\left[  1+O(x^{2})\right]  ,\;\,\mu=\pi/2,
\end{array}
\right.  ,\;0<\varkappa<1
\end{array}
\right.  ,\;x\rightarrow0.\label{4.1.7}%
\end{equation}

For the respective functions $h(\mu,s;x)=\phi^{\prime}(\mu,s;x)/\phi(\mu
,s;x)$, we have%

\begin{align}
h(\mu,s;x)  &  =\frac{1}{2x}+sk_{0}\frac{\sin\mu\,\Gamma(1+\varkappa)\left(
\frac{s}{2}\right)  ^{-\varkappa}I_{\varkappa}^{\prime}(sk_{0}x)+\cos
\mu\,\frac{2}{\Gamma(\varkappa)}\left(  \frac{s}{2}\right)  ^{\varkappa
}K_{\varkappa}^{\prime}(sk_{0}x)}{\sin\mu\,\Gamma(1+\varkappa)\left(
\frac{s}{2}\right)  ^{-\varkappa}I_{\varkappa}(sk_{0}x)+\cos\mu\,\frac{2}%
{\Gamma(\varkappa)}\left(  \frac{s}{2}\right)  ^{\varkappa}K_{\varkappa
}(sk_{0}x)},\nonumber\\
h(\mu,s;x)  &  =\left\{
\begin{array}
[c]{l}%
sk_{0}+O(x^{-1}),\,\,\mu\in\lbrack0,\pi/2)\\
-sk_{0}+O(x^{-1}),\,\mu=\pi/2
\end{array}
\right.  ,\;x\rightarrow\infty,\label{4.1.8}%
\end{align}
the prime $\prime\,$\ in the r.h.s. of (\ref{4.1.8})\ denotes a derivative
with respect to the argument.

\subsection{Initial operators $\hat{a}$ and $\hat{b}$}

We introduce the pairs of initial differential operators $\hat{a}(\mu,s)$ and
$\hat{b}(\mu,s)$ in $L^{2}(R_{+})$\textrm{\thinspace}defined on the subspace
$\mathcal{D}(\mathbb{R}_{+})$ of smooth functions compactly supported in
$(0,\infty)$, $D_{a(\mu,s)}=D_{b(\mu,s)}=\mathcal{D}(\mathbb{R}_{+})$, and
associated with each pair of the respective differential operations $\check
{a}(\mu,s)$ and $\check{b}(\mu,s)$ (\ref{4.1.4}). Similarly to the
differential operations $\check{a}$ and $\check{b}$, we often omit the
arguments $\mu$ and $s$ of operators\ $\hat{a}$ and $\hat{b}$ for breavity if
this does lead to misunderstanding, the arguments are written when needed.

These operators have the property%
\begin{equation}
\left(  \psi,\hat{a}\,\xi\right)  =\left(  \hat{b}\,\psi,\xi\right)
,\;\forall\psi,\xi\in\mathcal{D}(\mathbb{R}_{+}),\label{4.2.1}%
\end{equation}
which is easily verified by integration by parts.

It is evident from (\ref{4.1.5}) that the initial symmetric operator $\hat{H}
$ with $\alpha>-1/4\,$associated with $\check{H}$ and defined on
$\mathcal{D}(\mathbb{R}_{+})$ can be represented as%

\begin{equation}
\hat{H}=\hat{b}(\mu,s)\,\hat{a}(\mu,s)-(sk_{0})^{2}\hat{I}.\label{4.2.2}%
\end{equation}

These representations provide a basis for constructing generalized oscillator
representations for all s.a. Calogero Hamiltonians $\hat{H}_{\mathfrak{e}}$
with $\alpha>-1/4\,$in accordance with the program formulated in Introduction.
Namely, we should construct all possible extensions of any pair $\hat{a}(\mu
,s)$, $\hat{b}(\mu,s)$ of initial operators with given$\,\mu$ and $s$ to a
pair of closed mutually adjoint operators $\hat{c}(\mu,s) $, $\hat{c}^{+}%
(\mu,s)$, $\hat{a}(\mu,s)\subset\hat{c}(\mu,s)$, $\hat{b}(\mu,s)\subset
\hat{c}^{+}(\mu,s)$, or equivalently, to a pair of closed mutually adjoint
operators $\hat{d}(\mu,s)$, $\hat{d}^{+}(\mu,s)$, $\hat{a}(\mu,s)\subset
\hat{d}^{+}(\mu,s)$, $\hat{b}(\mu,s)\subset\hat{d}(\mu,s)(\mu,s)$. These
extensions produce the respective operators
\begin{equation}
\hat{H}_{\mathfrak{e\,}c(\mu,s)}=\hat{c}^{+}(\mu,s)\,\hat{c}(\mu
,s)-(sk_{0})^{2}\hat{I},\label{4.2.3}%
\end{equation}
and
\begin{equation}
\hat{H}_{\mathfrak{e\,}d(\mu,s)}=\hat{d}(\mu,s)\,\hat{d}^{+}(\mu
,s)-(sk_{0})^{2}\hat{I},\label{4.2.4}%
\end{equation}
which are certain Calogero Hamiltonians represented in a generalized
oscillator form. The both $\hat{H}_{\mathfrak{e\,}c(\mu,s)}$ and
$\hat{H}_{\mathfrak{e\,}d(\mu,s)}$ are bounded from below, in particular,
their spectra are bounded from below by $-(sk_{0})^{2}$. If $\ker
\hat{c}(\mu,s)\neq\{0\}$, then $\ker\hat{c}(\mu,s)$ is the ground eigenspace
(eigenstate ) of $\hat{H}_{\mathfrak{e\,}c(\mu,s)}$, while $E_{0}%
=-(sk_{0})^{2}$ is its ground-state energy; similarly, if $\ker\hat{d}^{+}%
(\mu,s)\neq\{0\}$, then $\ker\hat{d}^{+}(\mu,s)$ is the ground eigenspace
(eigenstate ) of $\hat{H}_{\mathfrak{e\,}d(\mu,s)}$, while $E_{0}%
=-(sk_{0})^{2}$ is its ground-state energy. It then remains to identify
$\hat{H}_{\mathfrak{e\,}c(\mu,s)}$ and $\hat{H}_{\mathfrak{e\,}d(\mu,s)}$ with
the known Calogero Hamiltonians and to hope that varying $\mu$ and $s$, we can
represent all Calogero Hamiltonians in the generalized oscillator form.

We continue with extending an arbitrary pair of initial operators
$\hat{a}(\mu,s)$, $\hat{b}(\mu,s)$ to a pair of closed mutually adjoint operators.

\subsection{Adjoint operators $\hat{a}^{+}$ and $\hat{b}^{+}$, closed
operators $\overline{\hat{a}}$ and $\overline{\hat{b}}$}

In this subsection, the arguments $\mu$ and $s$ of all operators involved are
omitted for brevity, they are implicitly implied.

Because the operators $\hat{a}$ and $\hat{b}$ are densely defined, they have
the adjoints, the respective $\hat{a}^{+}$ and $\hat{b}^{+}$. The defining
equation for $\hat{a}^{+}$, i.e., the equation for pairs $\psi\in D_{a^{+}}$
and $\eta=\hat{a}^{+}\psi$ forming the graph of the operator $\hat{a}^{+}$,
see \cite{GitTyV12}, reads%
\[
(\psi,\hat{a}\xi)=(\eta,\xi),\ \ \forall\xi\in\mathcal{D}(\mathbb{R}_{+}).
\]
The equality (\ref{4.2.1}) then implies that $\hat{b}$ $\subset\hat{a}^{+}$.
It follows that $\hat{a}^{+}$ is densely defined and in turn\textrm{\ }has the
adjoint $(\hat{a}^{+})^{+}$, while the operator $\hat{a}$ has a closure
$\overline{\hat{a}}=(\hat{a}^{+})^{+}\subseteq\hat{b}^{+}$ and $\left(
\overline{\hat{a}}\right)  ^{+}=\hat{a}^{+}$. Similarly, we obtain that
$\hat{a}\subset\hat{b}^{+}$, and therefore, the operator $\hat{b}$ has a
closure $\overline{\hat{b}}=(\hat{b}^{+})^{+}\subseteq\hat{a}^{+}$ and
$\left(  \overline{\hat{b}}\right)  ^{+}=\hat{b}_{{}}^{+}$. We thus
have\textrm{\ }the chains of inclusions%
\begin{equation}
\hat{a}\subset\overline{\hat{a}}=(\hat{a}^{+})^{+}\subseteq\hat{b}^{+}%
,\,\hat{b}\subset\overline{\hat{b}}\,=\,(\hat{b}^{+})^{+}\subseteq
\hat{a}^{+}.\label{4.3.2}%
\end{equation}

\subsection{Domains of operators $\hat{a}^{+}$, $\hat{b}^{+}$, $\overline
{\hat{a}}$ and $\overline{\hat{b}}$}

As a rule, in this subsection, the arguments $\mu$ and $s$ of all operators
involved, as well as of differential expressions $\check{a}$ and $\check{b}$
and of functions $\phi$ and $h$, are omitted for brevity, they are written
when needed.

We now describe the operators $\hat{a}^{+},\ \hat{b}^{+},\ \overline
{\hat{a}},$ and$\ \overline{\hat{b}}$.\ In evaluating these operators, we
follow \cite{AkhGl81,Naima69,GitTyV12} where one of the basics is the notion
of the natural domain\footnote{For s.a. differential operations $\check{f}=$
$\check{f}^{\ast}$, the natural domain is conventionally denoted by
$D_{\check{f}}^{\ast}$ for special reason, see \cite{Naima69,AkhGl81,GitTyV12}%
.} $D_{\check{f}}^{n}$\textrm{\ }$\subset L^{2}(\mathbb{R}_{+})$\textrm{\ }for
a given differential operation $\check{f}$\textrm{\ }\ that is the subspace of
square-integrable functions\textrm{\ }$\psi(x)$\textrm{\ }such that the
function $\check{f}$\textrm{\ }$\psi(x)$ has a sense\footnote{For example, if
$\check{f}$ is a differential operation of order $n$ with smooth coefficients,
$\psi$ is absolutely continuous together with its $n-1$ derivatives.} and is
also square integrable, $D_{\check{f}}^{n}$\textrm{\ }$=\{\psi(x):\psi
,\check{f}\mathrm{\ }\psi\in L^{2}(R_{+})\}$. The natural domain is the domain
of the maximum operator associated with a given differential operation.

We outline the results of the evaluation. The operators $\hat{a}^{+}$ and
$\overline{\hat{b}}$ are associated with the differential operation $\check
{b}=\check{a}^{\ast}$, while the operators $\hat{b}^{+}$ and $\ \overline
{\hat{a}}$\textrm{\ }are associated with the differential operation $\check
{a}=\check{b}^{\ast}$. We therefore dwell on the domains of the operators
involved, which either coincide with or belong to the natural domains for the
respective differential operations .

i) The domain of the operator $\hat{a}^{+}$ is the natural domain
for\footnote{The symbol ``a.c.'' is a contraction of ``absolutely
continuous''.} $\check{b}$:%
\begin{align}
&  D_{a^{+}}=D_{\check{b}}^{n}=\{\psi(x):\psi\;\text{\textrm{is a.c. in}
}\mathbb{R}_{+};\,\,\nonumber\\
&  \psi,\check{b}\psi=-\frac{1}{\phi(x)}\frac{d}{dx}\left(  \phi
(x)\psi(x)\right)  =-\psi^{\prime}-h\psi=\eta\in L^{2}(\mathbb{R}%
_{+})\}.\label{4.4.1}%
\end{align}
Because the functions $h$ are bounded at infinity, see (\ref{4.1.8}), it is
immediately established in a \ standard way that the functions $\psi$
belonging to $D_{\check{b}}^{n}\,$vanish at infinity,%
\begin{equation}
\psi(x)\rightarrow0,\,x\rightarrow\infty,\,\forall\psi\in D_{\check{b}}%
^{n}.\label{4.4.2}%
\end{equation}

A generic function $\psi$ belonging to $D_{\check{b}}^{n}$ can be considered
as the general solution of the inhomogeneous differential equation%
\begin{equation}
\check{b}\psi(x)=\eta(x)
\end{equation}
under the additional conditions that the both $\psi(x)$ and $\eta(x)$ are
square integrable on $\mathbb{R}_{+}$. Therefore, with taking (\ref{4.1.6})
and (\ref{4.1.7})\ into account, $\psi$ belonging to $D_{\check{b}(\mu,s)}%
^{n}$ allows the representation%

\begin{align}
&  \psi(x)=\frac{1}{\phi(\mu,s;x)}\left[  C-\int_{x_{0}}^{x}dy\,\phi
(\mu,s;y)\eta(y)\right]  ,\,\,\eta(x)=\check{b}\psi(x)\in L^{2}(\mathbb{R}%
_{+}),\nonumber\\
&  x_{0}=0\,\text{\textrm{for} }0<\varkappa<1\text{\textrm{and for} }\mu
=\pi/2,\,\varkappa\geq1;\,x_{0}>0\mathrm{\ }\text{\textrm{for}}\mathrm{\ }%
\mu\in\lbrack0,\pi/2),\varkappa\geq1,\nonumber\\
&  C\,\mathrm{is\ arbitrary\,const\,for\,}\mu\in\lbrack0,\pi
/2);\,C=0~\text{\textrm{for }}\mu=\pi/2.\label{4.4.4}%
\end{align}

Estimating the integral term in (\ref{4.4.4}) with the
Cauchy--Bunyakovskii\textrm{\ }inequality, we obtain that the asymptotic
behavior of functions $\psi\in D_{\check{b}}^{n}$ at the origin is given by

\textrm{\ }%
\begin{equation}
\psi(x)=\left\{
\begin{array}
[c]{l}%
O(x^{1/2}),\,\,\varkappa>1\\
\left\{
\begin{array}
[c]{l}%
O(x^{1/2}\sqrt{\ln\frac{1}{x}}),\,\,\mu\in\lbrack0,\pi/2)\\
O(x^{1/2}),\,\,\mu=\pi/2\,
\end{array}
,\,\varkappa=1\right. \\
\left\{
\begin{array}
[c]{l}%
\frac{C}{\cos\mu}(k_{0}x)^{\varkappa-1/2}[1+O(x^{2\varkappa})]+O(x^{1/2}%
),\,\,\mu\in\lbrack0,\pi/2)\\
O(x^{1/2}),\,\,\mu=\pi/2\,
\end{array}
,\,0<\varkappa<1\right.
\end{array}
\right.  ,\,x\rightarrow0.\label{4.4.5}%
\end{equation}

We note that for $0<\varkappa<1$, the natural domain $D_{\check{b}(\mu,s)}%
^{n}$ for $\check{b}(\mu,s)$ with $\mu\in\lbrack0,\pi/2)$ can be represented
as a direct sum of the form\textrm{\ }%
\[
D_{\check{b}(\mu,s)}^{n}=\{C\psi_{0}(\mu,s)\}+\tilde{D}_{\check{b}(\mu,s)}%
^{n},\,\,\mu\in\lbrack0,\pi/2),\,\,0<\varkappa<1,
\]
where the function $\psi_{0}$ $(\mu,s;x)\,$belonging to $D_{\check{b}(\mu
,s)}^{n}$ is given by%
\[
\psi_{0}(\mu,s;x)=\frac{1}{\phi(\mu,s;x)}\zeta\left(  x\right)  ,\mathrm{\,}%
\text{\textrm{so that}}\ \check{b}(\mu,s)\psi_{0}(\mu,s;x)=-\frac{1}{\phi
(\mu,s;x)}\zeta^{\prime}\left(  x\right)  ,
\]
$\zeta\left(  x\right)  $ is a fixed smooth function equal to\textrm{\ }$1$ in
a neighborhood of the origin and equal to $0$ for $x\geq x_{\infty}>0$, and
$\tilde{D}_{\check{b}(\mu,s)}^{n}$ is the subspace of functions belonging to
$D_{\check{b}(\mu,s)}^{n}$ and vanishing at the origin:%
\begin{equation}
\tilde{D}_{\check{b}(\mu,s)}^{n}=\left\{  \psi(x)\in D_{\check{b}(\mu,s)}%
^{n}:\,\psi(x)=O(x^{1/2}),\,x\rightarrow0\right\}  ,\,\mu\in\lbrack
0,\pi/2),\,0<\ \varkappa<1.\label{4.4.8}%
\end{equation}

The final result is given by
\begin{equation}
D_{a^{+}(\mu,s)}=\left\{
\begin{array}
[c]{l}%
D_{\check{b}(\mu,s)}^{n},\,\mu\in\lbrack0,\pi/2),\,\,\varkappa\geq
1,\,\text{\textrm{and}}\mathrm{\,}\,\mu=\pi/2\mathrm{\ },\,\varkappa>0,\\
\,D_{\check{b}(\mu,s)}^{n}=\{C\psi_{0}(\mu,s)\}+\tilde{D}_{\check{b}(\mu
,s)}^{n},\,\mu\in\lbrack0,\pi/2),\;0<\varkappa<1,
\end{array}
\right.  .\label{4.4.9}%
\end{equation}

with $D_{\check{b}(\mu,s)}^{n}$ given by (\ref{4.4.1}) and with estimates
(\ref{4.4.2}) at infinity and (\ref{4.4.5}) at the origin.

ii) The operator $\hat{b}^{+}$ is described quite similarly. Its domain is the
natural domain for $\check{a}=\check{b}^{\ast}$:
\begin{equation}
D_{b^{+}}=D_{\check{a}}^{n}=\{\chi(x):\chi\;\text{\textrm{is a.c. in}
}\mathbb{R}_{+};\,\chi,\,\check{a}\chi=\phi d_{x}\left(  \frac{1}{\phi}%
\chi\right)  =\chi^{\prime}-h\chi=\eta\in L^{2}(\mathbb{R}_{+}%
)\}.\label{4.4.10}%
\end{equation}
By the same reasoning as for the case of functions $\psi$ belonging to
$D_{\check{b}}^{n}$, the functions $\chi$ belonging to$\,D_{\check{a}}^{n}$
vanish at infinity,
\begin{equation}
\chi(x)\rightarrow0,\,x\rightarrow\infty,\,\forall\chi\in D_{\check{a}}%
^{n}.\label{4.4.11}%
\end{equation}
Using arguments similar to those for a generic function $\psi$ belonging to
$D_{\check{b}(\mu,s)}^{n}$, with the natural interchange $\phi\leftrightarrow
1/\phi$, , we establish that a generic function $\chi$ belonging to
$D_{\check{a}{(\mu,s)}}^{n}$ allows the representation%

\begin{align}
&  \chi(x)=\phi(\mu,s;x)\left[  D+\int_{x_{0}}^{x}dy\frac{1}{\phi(\mu
,s;y)}\eta(y)\right]  ,\;\text{ }\eta(x)=\check{a}\chi(x)\in L^{2}%
(\mathbb{R}_{+}),\nonumber\\
&  x_{0}=0\ \mathrm{for\,}\mu\in\lbrack0,\pi/2);\,x_{0}>0\ \mathrm{for\,}%
\mu=\pi/2,\nonumber\\
&  D\text{ }\mathrm{is\,an\,arbitrary\,constant\,for}\,\mu=\pi/2,\,\varkappa
\geq1\mathrm{and\,for}\,0<\varkappa<1;\,D=0\,\mathrm{for\,}\mu\in\lbrack
0,\pi/2),\,\varkappa\geq1.\label{4.4.12}%
\end{align}

The asymptotic behavior of functions $\chi\in$ $D_{\check{a}(\mu,s)}^{n}$ at
the origin is given by%

\begin{equation}
\chi(x)=\left\{
\begin{array}
[c]{l}%
O(x^{1/2}),\text{\textrm{\ }}\varkappa\geq1\\
\left\{
\begin{array}
[c]{l}%
D\cos\mu(k_{0}x)^{1/2-\varkappa}+O(x^{1/2}),\ \ \mu\in\lbrack0,\pi/2),\,\\
O(x^{1/2}),\,\mu=\pi/2,\,
\end{array}
,\,0<\varkappa<1\right.
\end{array}
\right.  ,\,x\rightarrow0.\label{4.4.13}%
\end{equation}
We note that for $0<\varkappa<1$, the natural domain $D_{\check{a}(\mu,s)}%
^{n}$ for $\check{a}(\mu,s)$ with $\mu\in\lbrack0,\pi/2)$ can be represented
as a direct sum of the form\textrm{\ }
\begin{equation}
D_{\check{a}(\mu,s)}^{n}=\{D\chi_{0}(\mu,s)\}+\tilde{D}_{\check{a}(\mu,s)}%
^{n},\,\mu\in\lbrack0,\pi/2),\,0<\varkappa<1,\label{4.4.14}%
\end{equation}
where the function $\chi_{0}(\mu,s;x)\,$belonging to $D_{\check{a}(\mu,s)}%
^{n}$ is given by%
\[
\chi_{0}(\mu,s;x)=\phi(\mu,s;x)\zeta\left(  x\right)  ,\mathrm{\,}%
\text{\textrm{so that}}\ \ \check{a}(\mu,s)\chi_{0}(\mu,s;x)=\phi
(\mu,s;x)\zeta^{\prime}\left(  x\right)  ,
\]
$\zeta\left(  x\right)  $ is a fixed smooth function equal to $1$ in a
neighborhood of the origin and equal to $0$ for $x\geq x_{\infty}>0$, and
$\tilde{D}_{\check{a}}^{n}$ is the subspace of functions belonging to
$D_{\check{a}}^{n}$ and vanishing at the origin:%
\begin{equation}
\tilde{D}_{\check{a}(\mu,s)}^{n}=\left\{  \chi(x)\in D_{\check{a}(\mu,s)}%
^{n}:\,\chi(x)=O(x^{1/2}),\;x\rightarrow0\right\}  ,\,\mu\in\lbrack
0,\pi/2),\,\,0<\varkappa<1.\label{4.4.16}%
\end{equation}
The final result is given by
\begin{equation}
D_{b^{+}(\mu,s)}=\left\{
\begin{array}
[c]{l}%
D_{\check{a}(\mu,s)}^{n},\,,\mu\in\lbrack0,\pi/2)\;\varkappa\geq
1,\,\mathrm{and\,}\,\mu=\pi/2,\,\varkappa>0,\\
D_{\check{a}\,(\mu,s)}^{n}\,=\{D\chi_{0}(\mu,s)\}+\tilde{D}_{\check{a}(\mu
,s)}^{n},\,\mu\in\lbrack0,\pi/2),\;0<\varkappa<1,
\end{array}
\right.  .\label{4.4.17}%
\end{equation}
with $D_{\check{a}(\mu,s)}^{n}$ given by (\ref{4.4.10}) and with estimates
(\ref{4.4.11}) at infinity and (\ref{4.4.13}) at the origin.

iii) The operator $\overline{\hat{a}}$ is evaluated in accordance with
(\ref{4.3.2}),\textrm{\thinspace}$\overline{\hat{a}}=(\hat{a}^{+}%
)^{+}\subseteq\hat{b}^{+}$: as a restriction of $\hat{b}^{+}$, this operator
is associated with $\check{a}$ and its domain \textrm{\ }belongs to or
coincides with $D_{\check{a}}^{n}$, while the defining equation for
\textrm{\thinspace}$\overline{\hat{a}}$ as $(\hat{a}^{+})^{+}$,%
\[
(\chi,\hat{a}^{+}\psi)-(\overline{\hat{a}}\chi,\psi)=0,\,\chi\in
D_{\bar{a}},\text{\thinspace}\forall\psi\in D_{\check{b}}^{n},
\]
is reduced to the equation for $D_{\bar{a}}$, i.e., for the functions $\chi\in
D_{\bar{a}}\subseteq D_{\check{a}}^{n}$, of the form%
\begin{equation}
(\chi,\check{b}\psi)-(\check{a}\chi,\psi)=0,\,\chi\in D_{\bar{a}}\subseteq
D_{\check{a}}^{n},\,\forall\psi\in D_{\check{b}}^{n}.\label{4.4.19}%
\end{equation}
Integrating by parts in $(\check{a}\chi,\psi)$ and taking asymptotic estimates
(\ref{4.4.2}), (\ref{4.4.11}) and (\ref{4.4.5}), (\ref{4.4.13}) into account,
we establish that for $\mu\in\lbrack0,\pi/2),\,\,\varkappa\geq1 $and for
$\mu=\pi/2,\,\varkappa>0$, eq. (\ref{4.4.19}) holds identically for
all$\,\chi\in D_{\check{a}(\mu,s)}^{n}$ , while for $\mu\in\lbrack0,\pi/2),$
$\,0<\varkappa<1$, eq. (\ref{4.4.19})\thinspace is reduced to
\[
\overline{D}C=0,\text{ }\forall C,
\]
which requires that $D=0$.

We finally obtain that
\begin{align}
&  \overline{\hat{a}}(\mu,s)=\hat{b}^{+}(\mu,s),\,\,\mu\in\lbrack
0,\pi/2),\;\varkappa\geq1\text{\textrm{\ and\thinspace}}\,\mu=\pi
/2,\;\varkappa>0,\nonumber\\
&  \,\text{\textrm{in particular, }}D_{\bar{a}(\mu,s)}=D_{\check{a}(\mu
,s)}^{n},\label{4.4.21}%
\end{align}
and \textrm{\ }
\begin{equation}
\overline{\hat{a}}(\mu,s)\subset\hat{b}^{+}(\mu,s),\,D_{\bar{a}(\mu
,s)}=\,\tilde{D}_{\check{a}(\mu,s)}^{n}\;\text{\textrm{(\ref{4.4.16})},}%
\,\mu\in\lbrack0,\pi/2),\;0<\varkappa<1.\label{4.4.22}%
\end{equation}

iv) Quite similarly, we find
\begin{align}
&  \overline{\hat{b}}(\mu,s)=\hat{a}^{+}(\mu,s),\,\mu\in\lbrack0,\pi
/2),\;\varkappa\geq1\text{\textrm{\ and\thinspace}}\,\mu=\pi/2,\;\varkappa
>0,\nonumber\\
&  \,\text{\textrm{in particular, }}D_{\overline{b}(\mu,s)}=D_{\check{b}%
(\mu,s)}^{n},\text{ }\label{4.4.23}%
\end{align}
and
\begin{equation}
\overline{\hat{b}}(\mu,s)\subset\hat{a}_{{}}^{+}(\mu,s),\,D_{\overline{b}%
(\mu,s)}=\widetilde{D}_{\check{b}(\mu,s)}^{n}\mathrm{\,(\ref{4.4.8})},\text{
}\,\mu\in\lbrack0,\pi/2),\;0<\varkappa<1.\label{4.4.24}%
\end{equation}

We note that equality (\ref{4.4.23}) and inclusion (\ref{4.4.24}) directly
follow from the respective previous equality (\ref{4.4.21}) and inclusion
(\ref{4.4.22}) by taking the adjoints, and only the domain $D_{\overline
{b}(\mu,s)}$ in the last case has to be evaluated.

We thus show that\textrm{\ }each pair $\check{a}(\mu,s)$, $\check{b}(\mu,s) $
of mutually adjoint\ by Lagrange differential operations $\ $(\ref{4.1.4})
providing generalized oscillator representation (\ref{4.1.5}) for $\check{H}%
$\ (\ref{2.1}) with $\alpha>-1/4$\textrm{\ }$(\varkappa>0)$ generates a
unique\textrm{\ }pair $\overline{\hat{a}}(\mu,s)=\hat{b}^{+}(\mu,s)$,
$\ \hat{a}^{+}(\mu,s)=\overline{\hat{b}}(\mu,s)$ of closed mutually
adjoint\textrm{\thinspace} operators for $\,\mu\in\lbrack0,\pi/2),\,s\in
\lbrack0,\infty),\,\varkappa\geq1$ and for $\mu=\pi/2,\,s\in\lbrack
0,\infty),\,\varkappa>0$, while for $\mu\in\lbrack0,\pi/2)$,$\,s\in
\lbrack0,\infty)$, $0<\varkappa<1$, each pair $\check{a}(\mu,s)$, $\check
{b}(\mu,s)\,$generates two different pairs $\overline{\hat{a}}(\mu
,s),\ \hat{a}^{+}(\mu,s)$ and $\,\hat{b}^{+}(\mu,s)$, $\overline{\hat{b}}%
(\mu,s)$ of closed mutually adjoint\textrm{\thinspace} operators such that
$\overline{\hat{a}}(\mu,s)\subset$ $\hat{b}^{+}(\mu,s) $ and $\overline
{\hat{b}}(\mu,s)\subset$ $\hat{a}^{+}(\mu,s)$. The operators $\overline
{\hat{a}}(\mu,s)$\ and $\hat{b}^{+}(\mu,s)$ are extensions of the initial
operator $\hat{a}$, they are associated with $\check{a}$ , and their domains
are given by the respective (\ref{4.4.21}), \ref{4.4.22}) and (\ref{4.4.17}),
(\ref{4.4.10}). The operators $\overline{\hat{b}}(\mu,s)$\ and $\hat{a}^{+}%
(\mu,s) $\ are extensions of the initial operator $\hat{b}(\mu,s)$, they are
associated with $\check{b}$, and their domains are given by the respective
(\ref{4.4.23}), \ref{4.4.24}) and (\ref{4.4.9}), (\ref{4.4.1}).

It is easy to prove\textrm{\ }that there are no other pairs of closed mutually
adjoint operators that are extensions of each pair $\hat{a}(\mu,s),\,\hat
{b}(\mu,s)$. Indeed, let $\hat{g},\,\hat{g}^{+}$ be such a pair, then because
$\overline{\hat{a}}(\mu,s)$ and $\overline{\hat{b}}(\mu,s) $ are minimum
closed extensions of the respective $\hat{a}(\mu,s)$ and $\,\hat{b}(\mu,s)$,
we have%
\[
\hat{a}(\mu,s)\subset\overline{\hat{a}}(\mu,s)\subseteq\hat{g}=\overline
{\hat{g}}=(\,\hat{g}^{+})^{+},\,\hat{b}(\mu,s)\subset\overline{\hat{b}}%
\,(\mu,s)\subseteq\hat{g}^{+}.
\]
It follows by taking the adjoints of these inclusions that
\[
\hat{g}^{+}\subseteq\hat{a}^{+}(\mu,s),\,\hat{g}\subseteq\hat{b}^{+}(\mu,s),
\]
so that we finally have
\[
\overline{\hat{a}}(\mu,s)\subseteq\hat{g}\subseteq\hat{b}^{+}(\mu,s),
\]
in particular,
\[
D_{\overline{a}(\mu,s)}\subseteq D_{g}\subseteq D_{b^{+}(\mu,s)}.
\]
It then directly follows from (\ref{4.4.21}) that $\hat{g}=\overline
{\hat{a}}(\mu,s)=\hat{b}^{+}(\mu,s)$ and therefore $\hat{g}^{+}=\overline
{\hat{b}}(\mu,s)=\hat{a}^{+}(\mu,s)$ for $\mu\in\lbrack0,\pi/2),\,s\in
\lbrack0,\infty),\,\varkappa\geq1$ and for$\,\mu=\pi/2,\,s\in\lbrack
0,\infty),\,\varkappa>0$, while for\textrm{\ }$\mu\in\lbrack0,\pi
/2),s\in\lbrack0,\infty),0<\varkappa<1$, it follows from (\ref{4.4.22}),
(\ref{4.4.14}) that the domains $D_{b^{+}(\mu,s)}=D_{\check{a}(\mu,s)}^{n}$
and $D_{\overline{a}(\mu,s)}=\tilde{D}_{\check{a}(\mu,s)}^{n}$ differ by a
one-dimensional subspace, so that either $\hat{g}=\overline{\hat{a}}(\mu,s)$,
and therefore $\hat{g}^{+}=\hat{a}^{+}(\mu,s)$, or $\hat{g}=\hat{b}^{+}%
(\mu,s)$, and therefore $\hat{g}^{+}=\overline{\hat{b}}(\mu,s)$.

\section{Point $\alpha=-1/4$ ($\varkappa=0$)}

A consideration in this section is completely similar to that in the previous
section for the case of $\alpha>-1/4\;(\varkappa>0)$. A difference with the
previous case is that for $\alpha=-1/4$, the inclusion of the point $s=0$ in
the range of admissible $s$ and getting the known oscillator representation
for $\check{H}$, see \cite{GitTyV11}, from generalized oscillator
representations (\ref{2.3}) with $s>0$ in the limit $s\rightarrow0 $ calls for
special investigation.

We therefore distinguish the case of $s>0$ and the case of $s=0.$

\subsection{Region $s>0$}

\subsubsection{Generalized oscillator representations for $\check{H}$,
differential operations $\check{a}$ and $\check{b}$}

For the coupling constant $\alpha=-1/4$, we use form (\ref{2.15}) with
$\varkappa=0$ of the general solution of eq. (\ref{2.8}) with substitutions
$A\rightarrow A\sqrt{k_{0}}$, $B\rightarrow B\sqrt{k_{0}}$,%

\begin{equation}
\phi(s;x)=A\sqrt{k_{0}x}I_{0}(sk_{0}x)+B\sqrt{k_{0}x}K_{0}(sk_{0}%
x).\label{5.2.1.1}%
\end{equation}

Because $\sqrt{k_{0}x}I_{0}(sk_{0}x)$ and $\sqrt{k_{0}x}K_{0}(sk_{0}x)$ are
real-valued linearly independent solutions, a condition $\operatorname{Im}%
\phi(x)=0$ requires that $\operatorname{Im}A=\operatorname{Im}B=0$. The
function $I_{0}(sk_{0}x)$ monotonically increases from $1$\textrm{\ }at $x=0$
to infinity as $x\rightarrow\infty$, while the function $K_{0}(sk_{0}x)$
monotonically decreases from infinity at $x=0$ to $0$ as $x\rightarrow\infty$.
It follows that $\phi(s;x)$ (\ref{5.2.1.1}) is real valued and positive in
$(0,\infty)$ iff $A\geq0$, $B\geq0$, $A+B>0$. Once again, a common constant
positive factor in coefficients $A$ and $B$ is irrelevant from the standpoint
of generalized oscillator representation (\ref{2.3}), (\ref{2.12}) for
$\check{H}$, so that we can set $A=\sin\mu$, $B=\cos\mu,$\ $\mu\in\lbrack
0,\pi/2]$ without loss of generality.

As a result, we have a two-parameter family $\{\check{a}(\mu,s),\check{b}%
(\mu,s)\}$ of different pairs of mutually adjoint first-order differential
operations,\textrm{\ }%
\begin{align}
&  \check{a}(\mu,s)=\check{b}^{\ast}(\mu,s)=d_{x}-h(\mu,s;x)=\phi
(\mu,s;x)d_{x}\frac{1}{\phi(\mu,s;x)},\nonumber\\
&  \check{b}(\mu,s)=\check{a}^{\ast}(\mu,s)=-d_{x}-h(\mu,s;x)=-\frac{1}%
{\phi(\mu,s;x)}d_{x}\phi(\mu,s;x),\nonumber\\
&  h(\mu,s;x)=\frac{\phi^{\prime}(\mu,s;x)}{\phi(\mu,s;x)},\nonumber\\
&  \phi(\mu,s;x)=\sin\mu\sqrt{k_{0}x}I_{0}(sk_{0}x)+\cos\mu\sqrt{k_{0}x}%
K_{0}(sk_{0}x),\,\nonumber\\
&  \mu\in\lbrack0,\frac{\pi}{2}],\,s>0,\label{5.1.1.2}%
\end{align}
providing a two-parameter family of different generalized oscillator
representations (\ref{2.3}) for $\check{H}$ ( \ref{2.1}) with $\alpha=-1/4$,%
\begin{equation}
\check{H}=\check{b}(\mu,s)\check{a}(\mu,s)-(sk_{0})^{2}.\label{5.1.1.3}%
\end{equation}

As before, we indicate the ranges of parameters $\mu$ and $s$ in formulas to
follow only if they differ from the whole ranges, these are $[0,\pi/2]$ for
$\mu$ and $(0,\infty)$ for $s$. In the main resulting formulas, we indicate
the ranges of the both parameters.

For the asymptotic behavior of the functions $\phi(\mu,s;x)$ and the functions
$1/\phi(\mu,s;x)$ at the origin, which we need below, we have%
\begin{align}
\phi(\mu,s;x)  &  =\left\{
\begin{array}
[c]{l}%
\tilde{A}(\mu,s)\sqrt{\kappa_{0}x}-\cos\mu\sqrt{\kappa_{0}x}\ln(\kappa
_{0}x)+O(x^{5/2}\ln x)),\;\mu\in\lbrack0,\pi/2)\\
\sqrt{\kappa_{0}x}+O(x^{5/2}),\;\mu=\pi/2
\end{array}
\right.  ,\;x\rightarrow0,\nonumber\\
\tilde{A}(\mu,s)  &  =\sin\mu+\cos\mu\lbrack\psi(1)-\ln(s/2)]\label{5.1.1.4}%
\end{align}

and%
\[
\frac{1}{\phi(\mu,s;x)}=\left\{
\begin{array}
[c]{l}%
-\frac{1}{\cos\mu\sqrt{\kappa_{0}x}\ln(\kappa_{0}x)}\frac{1}{1-\tilde{A}(\mu
,s)/[\cos\mu\ln(\kappa_{0}x)]}+O(x^{3/2}/\ln x),\;\mu\in\lbrack0,\pi/2)\\
\frac{1}{\sqrt{\kappa_{0}x}}+O(x^{3/2}),\;\mu=\pi/2,
\end{array}
\right.  ,\;x\rightarrow0.
\]

For the respective functions $h(\mu,s;x)=\phi^{\prime}(\mu,s;x)/\phi(\mu
,s;x)$, we have%
\begin{align*}
h(\mu,s;x)  &  =\frac{1}{2x}+(sk_{0})\frac{\sin\mu\,I_{1}(sk_{0}x)-\cos\mu\,
K_{1}(sk_{0}x)}{\sin\mu\,\,I_{0}(sk_{0}x)+\cos\mu\,\,K_{0}(sk_{0}x)},\\
h(\mu,s;x)  &  =\left\{\begin{array}[c]{l}sk_{0}+
O(x^{-1}),\;\mu\in\lbrack0,\pi/2)\\-sk_{0}+O(x^{-1}),\;\mu=0,
\end{array} \right. ,\;x\rightarrow\infty.
\end{align*}

\subsubsection{Initial operators $\hat{a}$ and $\hat{b}$}

We introduce the initial differential operators $\hat{a}(\mu,s)$ and $\hat
{b}(\mu,s)$ associated with\textrm{\ }each pair of the respective differential
operations $\check{a}(\mu,s)$\textrm{\ }and $\check{b}(\mu,s)$\thinspace
(\ref{5.1.1.2})\textrm{\ } and defined on $\mathcal{D}(\mathbb{R}_{+})$. These
operators and the initial symmetric operator $\hat{H}$ with $\alpha=-1/4$ have
the properties that are copies of (\ref{4.2.1}) and (\ref{4.2.2}), with the
change $s\in\lbrack0,\infty)$ to $s\in(0,\infty)$, which provides a basis for
constructing generalized oscillator representations similar to (\ref{4.2.3}),
(\ref{4.2.4}) for all s.a. Calogero Hamiltonians $\hat{H}_{\mathfrak{e}}$ with
$\alpha=-1/4$ by constructing all possible extensions of any pair
$\hat{a}(\mu,s)$, $\hat{b}(\mu,s)$ of initial operators to a pair of closed
mutually adjoint operators.

A procedure for extending presented below is completely similar to that in the
previous section\textrm{\ }for the case of $\alpha>-1/4\,(\varkappa>0)$. Once
again, we often omit the arguments $\mu$ and $s$ of operators\ $\hat{a}$ and
$\hat{b}$ for brevity writing them when needed.

\subsubsection{Adjoint operators $\hat{a}^{+}$ and $\hat{b}^{+}$, closed
operators $\overline{\hat{a}}$ and $\overline{\hat{b}}$}

Using arguments similar to those in subsec.4.3, we prove that the operators
$\hat{a}$ and $\hat{b}$ have adjoints, \ the respective $\hat{a}^{+}$ and
$\hat{b}^{+}$, and closures, the respective $\overline{\hat{a}}$ and
$\overline{\hat{b}}$, which form the chains of inclusions similar to
(\ref{4.3.2}),%
\begin{equation}
\hat{a}\subset\overline{\hat{a}}=(\hat{a}^{+})^{+}\subseteq\hat{b}^{+}%
,\,\hat{b}\subset\overline{\hat{b}}\,=\,(\hat{b}^{+})^{+}\subseteq
\hat{a}^{+}.\label{5.1.3.1}%
\end{equation}

An evaluation of the operators $\hat{a}^{+},\,\hat{b}^{+},\,\ \overline
{\hat{a}},\,$and $\overline{\hat{b}}$ is completely similar to that in subsec.
4.4 for the case of\textrm{\ }$\alpha>-1/4$. The operators $\overline{\hat{b}%
}$ and $\hat{a}^{+}$ are associated with the differential operation $\check
{b}=\check{a}^{\ast}$, while the operators $\ \overline{\hat{a}}$ and $\hat
{b}^{+}$\textrm{\ }are associated with the differential operation $\check
{a}=\check{b}^{\ast}$.

i) The domain of the operator $\hat{a}^{+}$ is the natural domain for
$\check{b}$, $D_{a^{+}}=D_{\check{b}}^{n}$, given by a copy of\textrm{\ }%
(\ref{4.4.1}). A generic function $\psi$ belonging to $D_{\check{b}}^{n}$
allows the representation%

\begin{align*}
&  \psi(x)=\frac{1}{\phi(\mu,s;x)}\left[C-\int_{0}^{x}dy\,
\phi(\mu,s;y)\eta(y)\right],\,
\eta(x)=\check{b}\psi(x)\in L^{2}(\mathbb{R}_{+}),\\
&  C\text{ }\mathrm{is}\text{ }\mathrm{an}\text{ }\mathrm{arbitrary}
\text{ }\mathrm{constant}\,\mathrm{for}\text{ }\mu\in\lbrack0,\pi/2),
\,C=0\text{ }\mathrm{for}\mu=\pi/2.
\end{align*}
The asymptotic behavior of functions $\psi\in D_{\check{b}}^{n}$ at infinity
and at the origin is given by%
\begin{align}
\psi(x)  &  \rightarrow0,\,x\rightarrow\infty,\nonumber\\
\psi(x)  &  =\left\{
\begin{array}
[c]{l}%
-\frac{C}{\cos\mu}\frac{1}{\sqrt{k_{0}x}\ln(k_{0}x)}[1+O(\frac{1}{\ln(k_{0}%
x)})]+O(x^{1/2}),\,\;\mu\in\lbrack0,\pi/2)\\
O(x^{1/2}),\,\mu=\pi/2
\end{array}
\right.  ,\;x\rightarrow0.\label{5.1.3.3}%
\end{align}

The natural domain $D_{\check{b}(\mu,s)}^{n}$ for $\check{b}(\mu,s)$ with
$\mu\in\lbrack0,\pi/2)$ can be represented as a direct sum\textrm{\ }of the
form%
\[
D_{\check{b}(\mu,s)}^{n}=\{C\psi_{0}(\mu,s)\}+\tilde{D}_{\check{b}(\mu,s)}%
^{n},\,\mu\in\lbrack0,\pi/2),
\]
where the function $\psi_{0}$ $(\mu,s;x)\,$belonging to $D_{\check{b}(\mu
,s)}^{n}$ is given by%
\[
\psi_{0}(\mu,s;x)=\frac{1}{\phi(\mu,s;x)}\zeta\left(  x\right)  ,\mathrm{\,}%
\text{\textrm{so that}}\ \check{b}(\mu,s)\psi_{0}(\mu,s;x)=-\frac{1}{\phi
(\mu,s;x)}\zeta^{\prime}\left(  x\right)  ,
\]
$\zeta\left(  x\right)  $ is a fixed smooth function equal to $1$ in a
neighborhood of the origin and equal to $0$ for $x\geq x_{\infty}>0$, and
$\tilde{D}_{\check{b}(\mu,s)}^{n}$ is the subspace of functions belonging to
$D_{\check{b}(\mu,s)}^{n}$ and vanishing at the origin:%
\begin{equation}
\tilde{D}_{\check{b}(\mu,s)}^{n}=\left\{  \psi(x)\in D_{\check{b}(\mu,s)}%
^{n}:\,\psi(x)=O(x^{1/2}),\,x\rightarrow0\right\}  ,\,\mu\in\lbrack
0,\pi/2).\label{5.1.3.6}%
\end{equation}

The final result is given by
\begin{equation}
D_{a^{+}(\mu,s)}=\left\{
\begin{array}
[c]{l}%
D_{\check{b}(\mu,s)}^{n}=\{C\psi_{0}(\mu,s)\}+\tilde{D}_{\check{b}(\mu,s)}%
^{n},\,\mu\in\lbrack0,\pi/2),\\
D_{\check{b}(\mu,s)}^{n},\;\mu=\pi/2,
\end{array}
\right. \label{5.1.3.7}%
\end{equation}
with $D_{\check{b}(\mu,s)}^{n}$ given by a copy of (\ref{4.4.1}) and with
estimates (\ref{5.1.3.3}) at infinity and at the origin.

ii) The domain of the operator $\hat{b}^{+}$ is the natural domain for
$\check{a}$, $D_{b^{+}}=D_{\check{a}}^{n}$, given by a copy of (\ref{4.4.10}%
).\ A generic function $\chi$ belonging to $D_{\check{a}}^{n}$ allows the
representation\textrm{\ }%
\begin{align}
\chi(x)  &  =\phi(\mu,s;x)\left[  D+\int_{x_{0}}^{x}dy\frac{1}{\phi(\mu
,s;y)}\eta(y)\right]  ,\text{ }\,\eta(x)=\check{a}\chi(x)\in L^{2}%
(\mathbb{R}_{+}),\nonumber\\
x_{0}  &  =0\mathrm{\ for}\text{ }\mu\in\lbrack0,\pi/2),\,x_{0}%
>0\ \mathrm{for}\ \mu=\pi/2,\,D\text{ }\mathrm{is}\text{ }\mathrm{an}\text{
}\mathrm{arbitrary}\text{ }\mathrm{constan}\text{\textrm{t.}}\label{5.1.3.8}%
\end{align}

The asymptotic behavior of functions $\chi\in D_{\check{a}}^{n}$ at infinity
and at the origin is given by%
\begin{align}
\chi(x)  &  \rightarrow0,\,x\rightarrow\infty,\nonumber\\
\chi(x)  &  =\left\{
\begin{array}
[c]{l}%
-D\cos\mu\sqrt{k_{0}x}\ln(k_{0}x)+O(x^{1/2}\ln^{1/2}(\frac{1}{x})),\,\mu
\in\lbrack0,\pi/2)\\
O(x^{1/2}\ln^{1/2}(\frac{1}{x})),\text{ }\mu=\pi/2
\end{array}
\right.  ,\;x\rightarrow0.\label{5.1.3.9}%
\end{align}

The natural domain $D_{\check{a}(\mu,s)}^{n}$ for $\check{a}(\mu,s)$ with
$\mu\in\lbrack0,\pi/2)$ can be represented as a direct sum\textrm{\ }of the
form%
\[
D_{\check{a}(\mu,s)}^{n}=\{D\chi_{0}(\mu,s)\}+\tilde{D}_{\check{a}(\mu,s)}%
^{n},\,\mu\in\lbrack0,\pi/2),
\]
where the function$\chi_{0}(\mu,s;x)\,$belonging to $D_{\check{a}(\mu,s)}^{n}
$ is given by%
\[
\chi_{0}(\mu,s;x)=\phi(\mu,s;x)\zeta\left(  x\right)  ,\mathrm{\,}%
\text{\textrm{so that}}\ \ \check{a}(\mu,s)\chi_{0}(\mu,s;x)=\phi
(\mu,s;x)\zeta^{\prime}\left(  x\right)  ,
\]
$\zeta\left(  x\right)  $ is a fixed smooth function equal to $1$\textrm{\ }in
a neighborhood of the origin and equal to $0$ for $x\geq x_{\infty}>0$, and
$\tilde{D}_{\check{a}(\mu,s)}^{n}$ is the subspace of functions belonging to
$D_{\check{a}(\mu,s)}^{n}$ and vanishing at the origin:%
\begin{equation}
\tilde{D}_{\check{a}(\mu,s)}^{n}=\left\{  \psi(x)\in D_{\check{a}(\mu,s)}%
^{n}:\,\chi(x)=O(x^{1/2}\ln^{1/2}(\frac{1}{x})),\,x\rightarrow0\right\}
,\,\mu\in\lbrack0,\pi/2).\label{5.1.3.12}%
\end{equation}

The final result is given by
\begin{equation}
D_{b^{+}(\mu,s)}=\left\{
\begin{array}
[c]{l}%
D_{\check{a}(\mu,s)}^{n}=\{D\chi_{0}(\mu,s)\}+\tilde{D}_{\check{a}(\mu,s)}%
^{n},\,\mu\in\lbrack0,\pi/2),\\
D_{\check{a}(\mu,s)}^{n},\,\mu=\pi/2,
\end{array}
\right. \label{5.1.3.13}%
\end{equation}
with $D_{\check{a}(\mu,s)}^{n}$ given by a copy of (\ref{4.4.10}) and with
estimates (\ref{5.1.3.9}) at infinity and at the origin.

iii) The operator $\overline{\hat{a}}$ is evaluated in accordance with
(\ref{5.1.3.1}),\textrm{\ \thinspace}$\overline{\hat{a}}=(\hat{a}^{+}%
)^{+}\subseteq\hat{b}^{+}\mathrm{:}$ as a restriction of $\hat{b}^{+}$, this
operator is associated with $\check{a}$ and its domain \textrm{\ }belongs to
or coincides with $D_{\check{a}}^{n}$, while the defining equation for
\textrm{\thinspace}$\overline{\hat{a}}$ as $(\hat{a}^{+})^{+}$ is reduced to
the equation for its domain$\,D_{\bar{a}}\subseteq D_{\check{a}}^{n}$ of the
form%
\begin{equation}
(\chi,\check{b}\psi)-(\check{a}\chi,\psi)=0,\,\chi\in D_{\bar{a}}\subseteq
D_{\check{a}}^{n},\,\forall\psi\in D_{\check{b}}^{n}.\label{5.1.3.14}%
\end{equation}
Integrating by parts in $(\check{a}\chi,\psi)$ and taking asymptotic estimates
(\ref{5.1.3.3}) and (\ref{5.1.3.9}) into account, we establish that for
$\mu=\pi/2$, eq. (\ref{5.1.3.14}) holds identically for all$\,\chi\in
D_{\check{a}(\pi/2,s)}^{n}$ , while for $\mu\in\lbrack0,\pi/2)$, eq.
(\ref{5.1.3.14})\thinspace is reduced to
\[
\overline{D}C=0,\text{ }\forall C,
\]
which requires that $D=0$.

We finally obtain that%
\begin{equation}
\overline{\hat{a}}(\pi/2,s)=\hat{b}^{+}(\pi/2,s),\,\text{\textrm{in
particular, }}D_{\bar{a}(\pi/2,s)}=D_{\check{a}(\pi/2,s)}^{n},\label{5.1.3.16}%
\end{equation}
and \textrm{\ }
\begin{equation}
\overline{\hat{a}}(\mu,s)\subset\hat{b}^{+}(\mu,s),\,D_{\bar{a}(\mu
,s)}=\,\tilde{D}_{\check{a}(\mu,s)}^{n}\;\mu\in\lbrack0,\pi
/2).\label{5.1.3.17}%
\end{equation}

iv) Quite similarly, we find%
\begin{equation}
\overline{\hat{b}}(\pi/2,s)=\hat{a}^{+}(\pi/2,s),\;\text{\textrm{in
particular, }}D_{\overline{b}(\pi/2,s)}=D_{\check{b}(\pi/2,s)}^{n},\text{
}\label{5.1.3.18}%
\end{equation}
and
\begin{equation}
\overline{\hat{b}}(\mu,s)\subset\hat{a}^{+}(\mu,s),\,D_{\overline{b}(\mu
,s)}=\widetilde{D}_{\check{b}(\mu,s)}^{n},\text{ }\mu\in\lbrack0,\pi
/2).\label{5.1.3.19}%
\end{equation}

We note that equality (\ref{5.1.3.18}) and inclusion (\ref{5.1.3.19}) directly
follow from the respective previous equality (\ref{5.1.3.16}) and inclusion
(\ref{5.1.3.17}) by taking the adjoints, and only the domain $D_{\overline
{b}(\mu,s)}$ in the last case has to be evaluated.

We thus show that\textrm{\ }each pair $\check{a}(\mu,s)$, $\check{b}(\mu,s) $
of mutually adjoint by Lagrange differential operations (\ref{5.1.1.2})
providing generalized oscillator representations (\ref{5.1.1.3}) for
$\check{H}$\ (\ref{2.1}) with $\alpha=-1/4$\textrm{\ }$(\varkappa=0)$
generates a unique\textrm{\ }pair $\overline{\hat{a}}(\pi/2,s)=\hat{b}^{+}%
(\pi/2,s)$, $\ \hat{a}^{+}(\pi/2,s)=\overline{\hat{b}}(\pi/2,s)$ of closed
mutually adjoint\textrm{\thinspace} operators for $\mu=\pi/2,\,s\in(0,\infty
)$, while for $\mu\in\lbrack0,\pi/2),\,s\in(0,\infty)$, each pair $\check
{a}(\mu,s)$, $\check{b}(\mu,s)\,$generates two different pairs $\overline
{\hat{a}}(\mu,s),\ \hat{a}^{+}(\mu,s)$ and $\,\hat{b}^{+}(\mu,s),\overline
{\hat{b}}(\mu,s)$ of closed mutually adjoint\textrm{\thinspace} operators such
that $\overline{\hat{a}}(\mu,s)\subset$ $\hat{b}^{+}(\mu,s)$ and
$\overline{\hat{b}}(\mu,s)\subset$ $\hat{a}^{+}(\mu,s)$. The operators
$\overline{\hat{a}}(\mu,s)$\ and $\hat{b}^{+}(\mu,s)$ are extensions of the
initial operator $\hat{a}$, they are associated with $\check{a}$ , and their
domains are given by the respective (\ref{5.1.3.16}), \ref{5.1.3.17}) and
(\ref{5.1.3.13}), the operators $\overline{\hat{b}}(\mu,s)$\ and
$\hat{a}^{+}(\mu,s)$\ are extensions of the initial operator $\hat{b}(\mu,s)$,
they are associated with $\check{b}$, and their domains are given by the
respective (\ref{5.1.3.18}), \ref{5.1.3.19}) and (\ref{5.1.3.7}).

Using arguments similar to those in subsec. 4.4, it is easy to prove that
there are no other pairs of closed mutually adjoint operators that are
extensions of each pair $\hat{a}(\mu,s),\,\hat{b}(\mu,s)$.

\subsection{Point $s=0$}

In this case, the general real-valued solution of eq. (\ref{2.8}) is given by%

\begin{align}
&  \phi(x)=A\sqrt{k_{0}x}+B\sqrt{k_{0}x}\ln(k_{0}x)=\left\{
\begin{array}
[c]{c}%
B\sqrt{k_{0}x}\ln(\sigma k_{0}x),\;B\neq0\\
A\sqrt{k_{0}x},\;B=0
\end{array}
\right.  ,\label{5.2.1.1a}\\
&  \sigma=e^{A/B},\;\operatorname{Im}A=\operatorname{Im}B=0.\nonumber
\end{align}

It is evident that the real-valued $\phi(x)$ (\ref{5.2.1.1a}) is positive in
$(0,\infty)$ iff $A>0$, $B=0$. Because a constant positive factor is
irrelevant from the standpoint of generalized oscillator representation
(\ref{2.3}), (\ref{2.12}) for $\check{H}$, we can set $A=1$without loss of generality.

As a result, we have a unique pair $\check{a},\check{b}$ of mutually adjoint
first-order differential operations,\textrm{\ }%
\begin{align}
&  \check{a}=\check{b}^{\ast}=d_{x}-\frac{1}{2x}=\phi(x)d_{x}\frac{1}{\phi
(x)},\nonumber\\
&  \check{b}=\check{a}^{\ast}=-d_{x}-\frac{1}{2x}=-\frac{1}{\phi(x)}d_{x}%
\phi(x),\nonumber\\
&  h(\mu,s;x)=\frac{\phi^{\prime}(x)}{\phi(x)}=\frac{1}{2x},\;\phi
(x)=\sqrt{k_{0}x},\label{5.2.1.2}%
\end{align}
providing unique oscillator representation (\ref{2.3}) with $s=0$ for
$\check{H}$ with $\alpha=-1/4$,%
\begin{equation}
\check{H}=\check{b}\check{a}.\label{5.2.1.3}%
\end{equation}

We then introduce the initial operators $\hat{a}$ and $\hat{b}$ associated
with the respective $\check{a}$ and $\check{b}$ and construct the pairs of
operators $\overline{\hat{a}}$, $\hat{a}^{+}$ and $\overline{\hat{b}}$,
$\hat{b}^{+}$ as all possible extensions of the pair $\hat{a}$, $\hat{b}$ to a
pair of mutually adjoint closed operators. A procedure follows the standard
way adopted in the previous subsection. The result can be formulated as follows.

It is easy to see that formulas (\ref{5.2.1.2}) and (\ref{5.2.1.3}) can be
obtained from formulas (\ref{5.1.1.2}) and (\ref{5.1.1.3}) by setting $\mu
=\pi/2$ and taking the limit $s\rightarrow0$. Moreover, we can verify that all
the results concerning the properties of operators $\hat{a}^{+}$, $\hat{b}%
^{+}$, $\overline{\hat{a}}$, $\overline{\hat{b}}$, including their domains and
the equalities $\overline{\hat{a}}=\hat{b}^{+}$, $\overline{\hat{b}%
}=\hat{a}^{+}$, can be obtained from the corresponding results for the
operators $\hat{a}^{+}(\mu,s)$, $\hat{b}^{+}(\mu,s)$, $\overline{\hat{a}}%
(\mu,s)$, $\overline{\hat{b}}(\mu,s)$ in the preceding subsection by setting
$\mu=\pi/2$ and taking the limit $s\rightarrow0$, so that we can set
$\hat{a}^{+}=\hat{a}^{+}(\pi/2,0)$, $\hat{b}^{+}=\hat{b}^{+}(\pi/2,0)$,
$\overline{\hat{a}}=\overline{\hat{a}}(\pi/2,0)$, $\overline{\hat{b}%
}=\overline{\hat{b}}(\pi/2,0)$.

The final conclusion of this section on the case of the coupling constant
$\alpha=-1/4$ is that the results of the previous subsection for the operators
$\hat{a}^{+}(\mu,s)$, $\hat{b}^{+}(\mu,s)$, $\overline{\hat{a}}(\mu,s)$,
$\overline{\hat{b}}(\mu,s)$, where $\mu\in\lbrack0,\pi/2]$, $s\in(0,\infty)$,
actually hold for $\mu\in\lbrack0,\pi/2)$, $s\in(0,\infty)$ and for $\mu
=\pi/2$ , $s\in\lbrack0,\infty)$.

\section{Oscillator representations}

Now we are in a position to answer the question on generalized oscillator
representations%
\begin{equation}
\hat{H}_{\mathfrak{e}}=\hat{c}^{+}\hat{c}-(sk_{0})^{2}\hat{I},\,s\geq
0,\label{6.1a}%
\end{equation}
or equivalently
\begin{equation}
\hat{H}_{\mathfrak{e}}=\hat{d}\,\hat{d}^{+}-(sk_{0})^{2}\hat{I},\,s\geq
0,\label{6.1b}%
\end{equation}
where $\hat{c}$ and $\hat{d}$ are densely defined closed first-order
differential operators and $\hat{c}^{+}$ and $\hat{d}^{+}$\ are their
respective adjoints, for all Calogero Hamiltonians $\hat{H}_{\mathfrak{e}}$
with any coupling constant $\alpha\in(-\infty,\infty)$. We recall that any
$\hat{H}_{\mathfrak{e}}$ is a s.a. operator associated with Calogero
differential operation $\check{H}$ (\ref{2.1}) with the same $\alpha$. An
answer to the question is essentially different for the region $\alpha<-1/4$
and for the region $\alpha\geq-1/4$.

We can say immediately that any Calogero Hamiltonian $\hat{H}_{\mathfrak{e}}$
with $\alpha<-1/4$ does not allow generalized oscillator representations
(\ref{6.1a}) or (\ref{6.1b}) because such a representation would imply that
$\hat{H}_{\mathfrak{e}}$ is bounded from below, whereas any Calogero
Hamiltonian with $\alpha<-1/4$ is not bounded from below \cite{GitTyV12}.
This\ conclusion is in complete agreement\ with that according to sec. 3,
there is no generalized oscillator representation (\ref{2.3}), (\ref{2.12})
for Calogero differential operation $\check{H}$ (\ref{2.1}) with $\alpha<-1/4
$.

As to the second region of the couling constant $\alpha\geq-1/4$, we show in
what follows that any Calogero Hamiltonian with $\alpha\geq1/4$ does allow
generalized oscillator representations (\ref{6.1a}) or (\ref{6.1b}), in fact,
a family of such representations, one- or two-parameter. In accordance with
the program formulated in Introduction and according to the above results, we
have two families of Calogero Hamiltonians in a generalized oscillator form,
\begin{equation}
\hat{H}_{\mathfrak{e}a(\mu,s)}=\hat{a}^{+}(\mu,s)\overline{\hat{a}}%
(\mu,s)-(sk_{0})^{2}\hat{I},\,\mu\in\lbrack0,\pi/2],\;s\in\lbrack
0,\infty),\label{6.2}%
\end{equation}
and
\begin{equation}
\hat{H}_{\mathfrak{e}b(\mu,s)}=\overline{\hat{b}\ }(\mu,s)\hat{b}^{+}%
(\mu,s)-(sk_{0})^{2}\hat{I},\mu\in\lbrack0,\pi/2],\;s\in\lbrack0,\infty
),\label{6.3}%
\end{equation}
for any $\alpha\geq-1/4$. It turns out that these families cover all the set
of the known Calogero Hamiltonians with given $\alpha\geq-1/4$. Namely, each
Calogero Hamiltonian with given $\alpha\geq-1/4$ can be identified with one or
more members of family (\ref{6.2}) or family (\ref{6.3}). This identification
is trivial in the case of $\alpha\geq3/4$ where there is a unique Calogero
Hamiltonian with given $\alpha$. In the case of $\alpha$ such that
$-1/4\leq\alpha$ $<3/4$, the procedure of identification is more complicated.
We recall that for each $\alpha\in\lbrack-1/4,3/4)$, there exists a
one-parameter family $\{\hat{H}_{\nu},$ $\nu\in\lbrack-\pi/2,\pi/2]\}$ of
Calogero Hamiltonians differing in their domains $D_{H_{\nu}}$\cite{GitTyV12}.
Namely, the domains $D_{H_{\nu}}$ are subspaces of the natural domain
$D_{\check{H}}^{n}$ for $\check{H}$ that are specified by different s.a.
asymptotic boundary conditions at the origin..Therefore, the identification of
a given $\hat{H}_{\mathfrak{e}a(\mu,s)}$ with a certain $\hat{H}_{\nu}$ goes
through evaluating the asymptotic behavior of functions belonging to the
domain of $\hat{H}_{\mathfrak{e}a(\mu,s)}$ and its identification with the
boundary conditions for this $\hat{H}_{\nu}$; the same holds for
$\hat{H}_{\mathfrak{e}b(\mu,s)}$. It may happen, and that really occurs, that
different $\hat{H}_{\mathfrak{e}a(\mu,s)}$ or $\hat{H}_{\mathfrak{e}b(\mu,s)}$
have the same asymptotic behavior of functions belonging to their domains.

We first consider the Hamiltonians $\hat{H}_{\mathfrak{e}a}$.

\subsection{Hamiltonians $\hat{H}_{\mathfrak{e}a}$}

\subsubsection{Region $\alpha\geq3/4$ ($\varkappa\geq1$)}

For each $\alpha$ in this region, there exists only one s.a. Calogero
Hamiltonian $\hat{H}_{1}$ defined on the\textrm{\ }natural domain
$D_{\check{H}}^{n}$ , see. \cite{GitTyV12}. That is why we can immediately
conclude that%
\begin{equation}
\hat{H}_{1}=\hat{a}^{+}(\mu,s)\overline{\hat{a}}(\mu,s)-(sk_{0})^{2}%
\hat{I},\ \forall\mu\in\lbrack0,\pi/2],\,\forall s\in\lbrack0,\infty
),\,\varkappa\geq1,\label{6.1.1.1}%
\end{equation}
which represents a two-parameter family of oscillator representations for a
unique Calogero Hamiltonian $\hat{H}_{1}$ with given coupling constant
$\alpha\geq3/4$.

Taking\textrm{\ }$s=0$\textrm{\ }in the r.h.s of (\ref{6.1.1.1}),
\textrm{\ }we obtain the known one-parameter family of oscillator
representations for the nonnegative $\hat{H}_{1}$ \cite{GitTyV11} which are
the optimum representations from the standpoint of an optimum estimate on its
spectrum from below.

According to \cite{GitTyV12}, the spectrum of $\hat{H}_{1}$ is given by
$\mathrm{spec}\hat{H}_{1}=[0,\infty)$ and is continuous, which agrees with
that $\ker\overline{\hat{a}}(\mu,s)=\{0\}$, $\forall\mu,\,\forall s$.

\subsubsection{Region $-1/4<\alpha<3/4$ ($0<\varkappa<1$)}

By definition of the operator $\hat{H}_{\mathfrak{e}a(\mu,s)}$ , its domain
belongs to or coincides with the domain of the operator\textrm{\ }%
$\overline{\hat{a}}(\mu,s)$,\textrm{\ }$D_{H_{\mathfrak{e}a(\mu,s)}}\subseteq
D_{\bar{a}(\mu,s)}$\textrm{\ }given by (\ref{4.4.21}) for $\mu=\pi/2$ and
(\ref{4.4.22}) for $\mu\in\lbrack0,\pi/2)$. According to (\ref{4.4.13}) and
(\ref{4.4.16}), the asymptotic behavior of functions $\chi$ belonging to
$D_{\bar{a}(\mu,s)}$ is estimated for the case of $0<\varkappa<1$ by
$\chi=O(x^{1/2})$ as $x\rightarrow0$. It follows that the functions belonging
to $D_{H_{\mathfrak{e}a(\mu,s)}}$\textrm{\ }tend to zero not weaker than
$x^{1/2}$ as $x\rightarrow0$.

According to \cite{GitTyV12}, \textrm{t}here exists only one s.a. Calogero
Hamiltonian with such asymptotic behavior of functions belonging to its domain
at the origin, namely, $\hat{H}_{2,\pm\pi/2}$. We thus establish that
\begin{equation}
\hat{H}_{2,\pm\pi/2}=\hat{a}^{+}(\mu,s)\overline{\hat{a}}(\mu,s)-(sk_{0}%
)^{2}\hat{I},\ \forall\mu\in\lbrack0,\pi/2],\,\forall s\in\lbrack
0,\infty),\ 0<\varkappa<1,\label{6.1.2.1}%
\end{equation}
which represents a two-parameter family of oscillator representations for a
unique Calogero Hamiltonian $\hat{H}_{2,\pm\pi/2}$ with given coupling
constant $\alpha\in($ $-1/4<\alpha<3/4)$.

Taking\textrm{\ }$s=0$\textrm{\ }in the r.h.s of (\ref{6.1.2.1}),
\textrm{\ }we obtain the known one-parameter family of oscillator
representations for the nonnegative $\hat{H}_{2,\pm\pi/2}$ \cite{GitTyV11}
which are the optimum representations from the standpoint of an optimum
estimate on its spectrum from below.

According to \cite{GitTyV12}, the spectrum of $\hat{H}_{2,\pm\pi/2}$ is given
by $\mathrm{spec}\hat{H}_{2,\pm\pi/2}=[0,\infty)$ and is continuous, which
agrees with that $\ker\overline{\hat{a}}(\mu,s)=\{0\}$, $\forall\mu,\,\forall
s$.

\subsubsection{Point $\alpha=-1/4$ ($\varkappa=0$)}

A reasoning in this case is completely similar to that in the previous case of
$-1/4<\alpha<3/4$.

By definition of the operator $\hat{H}_{\mathfrak{e}a(\mu,s)}$, its domain
$D_{H_{\mathfrak{e}a}(\mu,s)}$ belongs to or coincides with the domain
$D_{\bar{a}(\mu,s)}$ given by (\ref{5.1.3.16}) for $\mu=\pi/2$, the point $s=0
$ included, see subsec. 5.2, and (\ref{5.1.3.17}) for $\mu\in\lbrack0,\pi/2)$.
According to (\ref{5.1.3.9}) and (\ref{5.1.3.12}), the asymptotic behavior of
functions $\chi$ belonging to $D_{\bar{a}(\mu,s)}$ is estimated by
$\chi=O(x^{1/2}\ln^{1/2}\frac{1}{x})$ as $x\rightarrow0$. It follows that the
functions belonging to $D_{H_{\mathfrak{e}a(\mu,s)}}$\textrm{\ }tend to zero
not weaker than $x^{1/2}\ln^{1/2}\frac{1}{x}$ as $x\rightarrow0$.

According to \cite{GitTyV12}, there exists only one s.a. Calogero Hamiltonian
with such asymptotic behavior of functions belonging to its domain at the
origin, namely, $\hat{H}_{3,\pm\pi/2}$. We thus establish that
\begin{equation}
\hat{H}_{3,\pm\pi/2}=\hat{a}^{+}(\mu,s)\overline{\hat{a}}(\mu,s)-(sk_{0}%
)^{2}\hat{I},\ \left\{
\begin{array}
[c]{l}%
\forall\mu\in\lbrack0,\pi/2),\,\forall s>0\\
\mu=\pi/2,\,\forall s\geq0
\end{array}
\right.  \ ,\,\varkappa=0,\label{6.1.3.1}%
\end{equation}
which represents a two-parameter family of oscillator representations for a
unique Calogero Hamiltonian $\hat{H}_{3,\pm\pi/2}$ with coupling constant
$\alpha=-1/4$.

Taking\textrm{\ }$\mu=\pi/2,\,s=0$ \textrm{\thinspace}in the r.h.s of
(\ref{6.1.3.1}), \textrm{\ }we obtain the known oscillator representation for
the nonnegative $\hat{H}_{3,\pm\pi/2}$ \cite{GitTyV11} which is the optimum
representation from the standpoint of an optimum estimate on its spectrum from below.

According to \cite{GitTyV12}, the spectrum of $\hat{H}_{3,\pm\pi/2}$ is given
by $\mathrm{spec}\hat{H}_{3,\pm\pi/2}=[0,\infty)$ and is continuous, which
agrees with that $\ker\overline{\hat{a}}(\mu,s)=\{0\}$, $\forall\mu,\,\forall
s$.

\subsection{Hamiltonians $\hat{H}_{\mathfrak{e}b}$}

\subsubsection{Region $\alpha\geq3/4$ ($\varkappa\geq1$)}

For each $\alpha$ in this region, we have the identities $\overline{\hat{b}%
}(\mu,s)=\hat{a}^{+}(\mu,s)$ and $\,\hat{b}^{+}(\mu,s)=\overline{\hat{a}}%
(\mu,s)$, see (\ref{4.4.23}) and (\ref{4.4.21}), so that with taking into
account subsubsec. 6.1.1, formula (\ref{6.1.1.1}), we find
\begin{equation}
\hat{H}_{1}=\overline{\hat{b}}(\mu,s)\hat{b}^{+}(\mu,s)-(sk_{0})^{2}%
\hat{I},\,\forall\mu\in\lbrack0,\pi/2],\;\forall s\geq0,\varkappa
\geq1,\label{6.2.1.1}%
\end{equation}
which is another representation of the known two-parameter family of
oscillator representations (\ref{6.1.1.1}) for $\hat{H}_{1}$ followed by an
appropriate comment.

\subsubsection{Region $-1/4<\alpha<3/4$ ($0<\varkappa<1$), $\mu=\pi/2$}

A reasoning concerning $\hat{H}_{\mathfrak{e}b}$ in this case is completely
similar to that in the previous subsubsection for the case of $\alpha\geq3/4$.
According to (\ref{4.4.23}) and (\ref{4.4.21}), we have the identities
$\overline{\hat{b}}(\pi/2,s)=\hat{a}^{+}(\pi/2,s)$ and $\,\hat{b}^{+}%
(\pi/2,s)=\overline{\hat{a}}(\pi/2,s)$, so that with taking into account
subsubsec. 6.1.2, formula (\ref{6.1.2.1}), we find
\begin{equation}
\hat{H}_{_{2,\pm\pi/2}}=\overline{\hat{b}}(\pi/2,s)\hat{b}^{+}(\pi
/2,s)-(sk_{0})^{2}\hat{I},\,\forall s\geq0,0<\varkappa<1,\label{6.2.2.1}%
\end{equation}
which is another representation of the one-parameter family of oscillator
representations for $\hat{H}_{_{2,\pm\pi/2}}$ that is a restriction of the
known two-parameter-family of oscillator representations (\ref{6.1.2.1}) to
$\mu=\pi/2$. Of course, \ the comment following (\ref{6.1.2.1}) holds.

\subsubsection{$-1/4<\alpha<3/4$ ($0<\varkappa<1$),\thinspace$0\leq\mu<\pi/2$}

We have to find the asymptotic behavior of functions belonging to the domain
$D_{H_{\mathfrak{e}b(\mu,s)}}$ of the operator $\hat{H}_{\mathfrak{e}b(\mu
,s)},$ $\mu\in\lbrack0,\pi/2),\,\,s\in\lbrack0,\infty),\,\,\varkappa\in(0,1)$,
at the origin.

We begin with representing the asymptotic behavior of functions $\phi
(\mu,s;x)$ (\ref{4.1.4}), $\mu\in\lbrack0,\pi/2)$,$\,\,s\in\lbrack0,\infty)$,
$\varkappa\in(0,1)$, at the origin.given in\textrm{\ }(\ref{4.1.6}) in a new form:%

\begin{align*}
&  \phi(\mu,s;x)=c[(k_{0}x)^{1/2+\varkappa}\sin\theta(\mu,s)+(k_{0}%
x)^{1/2-\varkappa}\cos\theta(\mu,s)]+O(x^{5/2-\varkappa}),\,,x\rightarrow
0,\\ &  \tan\
theta(\mu,s)=\tan\mu-\frac{\Gamma(1-\varkappa)}{\Gamma(1+\varkappa)}\left(
\frac{s}{2}\right)  ^{2\varkappa},\,c=\frac{\cos\mu}{\cos\theta(\mu,s)},\\ &
\mu\in\lbrack0
,\frac{\pi}{2}),\,s\in\lbrack0,\infty)\,,\,\varkappa\in(0,1),\,\mathrm
{\,}\theta\in\left(  -\frac{\pi}{2},\frac{\pi}{2}\right)  .
\end{align*}
By definition of the operator $\hat{H}_{\mathfrak{e}b(\mu,s)}$, its domain
$D_{H_{\mathfrak{e}b(\mu,s)}}$ consists of functions $\chi\in D_{b^{+}(\mu
,s)}$ such that $\hat{b}^{+}(\mu,s)\chi=\eta\in D_{\overline{b}(\mu,s)}\subset
L^{2}(\mathbb{R}_{+})$. The first condition implies that $\chi$ allows
representation (\ref{4.4.12}) with $x_{0}=0$ and, in general, $D\neq0 $, while
the second condition implies that $\eta(x)=O(x^{1/2}),\;x\rightarrow0$, see
(\ref{4.4.24}),\textrm{\ }(\ref{4.4.8}). Estimating the integral term in
(\ref{4.4.12}) with such $\eta$, we obtain that the asymptotic behavior of
functions $\chi\in$ $D_{H_{\mathfrak{e}b(\mu,s)}}$, $\mu\in\lbrack
0,\pi/2),\,s\in\lbrack0,\infty)\,,\,\varkappa\in(0,1)$, at the origin is given
by%
\begin{equation}
\chi(x)=C[(k_{0}x)^{1/2+\varkappa}\sin\theta(\mu,s)+(k_{0}x)^{1/2-\varkappa
}\cos\theta(\mu,s)]+O(x^{3/2}),\,x\rightarrow0.\label{6.2.3.2}%
\end{equation}

According to \cite{GitTyV12}, for each $\alpha\in(-1/4,3/4)\,(\varkappa
\in(0,1))$, there is a one-parameter family of s.a. Calogero Hamiltonians with
such asymptotic behavior of functions belonging to their domains, namely , the
family $\{\hat{H}_{2,\nu},\,\nu\in\left(  -\frac{\pi}{2},\frac{\pi}{2}\right)
\}$. The parameter $\nu$ is naturally identified with the angle $\theta
(\mu,s)$ in (\ref{6.2.3.2}), and we establish that%

\begin{align}
&  \hat{H}_{2,\nu}=\overline{\hat{b}}(\mu,s)\hat{b}^{+}(\mu,s)-(sk_{0}%
)^{2},\text{ }\nu=.\theta(\mu,s),\nonumber\\
&  \nu\in(-\pi/2,\pi/2),\,\mu\in\lbrack0,\pi/2),\,s\in\lbrack0,\infty
),\,\varkappa\in(0,1),\label{6.2.3.3}%
\end{align}
which represents a one-parameter family\textrm{\ }of generalized oscillator
representations for $\hat{H}_{2,\nu}$ with given $\alpha\in
(-1/4,3/4),\,(0<\varkappa<1),$ and $\nu\in(-\pi/2,\pi/2)$.

It is convenient to take $\mu$ as the independent parameter, then $s$ is
easily determined from the relation $\tan\vartheta(\mu,s)=\tan\nu$ to yield%

\begin{align}
s &  =s(\mu,\nu)=2\left[  (\tan\mu-\tan\nu)\frac{\Gamma(1+\varkappa)}%
{\Gamma(1-\varkappa)}\right]  ^{1/2\varkappa},\nonumber\\
\nu &  \in(-\pi/2,\pi/2),\,\mu\in\lbrack0,\pi/2),\,\varkappa\in
(0,1),\label{6.2.3.4}%
\end{align}
with due regard to the condition $s\geq0$.$\ $For fixed $\nu$, the function
$s(\mu,\nu)$ (\ref{6.2.3.4}) is monotonically increasing from $s_{\min}(\nu)$
to $\infty$ as $\mu$ ranges admissible values from $\mu_{\min}(\nu)$ to
$\pi/2-0$, where%
\[
\left\{
\begin{array}
[c]{c}%
\mu_{\min}(\nu)=\nu,\;s_{\min}(\nu)=0,\;0\leq\nu<\pi/2,\\
\mu_{\min}(\nu)=0,\;s_{\min}(\nu)=2\left[  \tan|\nu|\frac{\Gamma(1+\varkappa
)}{\Gamma(1-\varkappa)}\right]  ^{1/2\varkappa},-\pi/2<\,\nu<0
\end{array}
\right.  .
\]
It is evident that the spectrum of $\hat{H}_{2,\nu}$ is bounded from below
by$\,-(s_{\min}(\nu)k_{0})^{2}$.

If $0\leq\nu<\pi/2$, this boundary is zero, and according to \cite{GitTyV12},
this is an exact lower boundary of the spectrum, \textrm{Spec}$\hat{H}_{2,\nu
}=[0,\infty)$, and the spectrum is continuous, which agrees with that
$\ker\hat{b}^{+}(\mu,s(\nu,\mu))=\{0\}$, $\forall\mu\in\lbrack\nu,\pi/2) $.

Taking $\mu=\mu_{\min}(\nu)=\nu$ and $s=s(\mu_{\min}(\nu),\nu)=s_{\min}%
(\nu)=0$, we obtain the known oscillator representation for the nonnegative
$\hat{H}_{2,\nu}$, $0\leq\nu<\pi/2$, \cite{GitTyV11}, which is the optimum representation.

If $-\pi/2<\nu<0$, we have $\ker\hat{b}^{+}(0,s_{\min}(\nu))=\{c\phi
(0,s_{\min}(\nu);x)\}\neq\{0\}$ because $\phi(0,s_{\min}(\nu);x)=K_{\varkappa
}(s_{\min}(\nu)k_{0}x)$ is square integrable on the whole semiaxis
$\mathbb{R}_{+}$, whereas $\ker\hat{b}^{+}(\mu,s(\mu,\nu))=\{0\}$,
$\,0<\mu<\pi/2$. This implies that%
\[
-(s_{\min}(\nu)k_{0})^{2}=-4k_{0}^{2}\left|  \tan\nu\frac{\Gamma(1+\varkappa
)}{\Gamma(1-\varkappa)}\right|  ^{1/\varkappa}=E_{2}(\nu)
\]
is an exact lower boundary of the spectrum of $\hat{H}_{2,\nu}$, $E_{2}(\nu)$
is an eigenvalue of $\hat{H}_{2,\nu}$, the energy of its negative ground
level, and the normalized eigenfunction of the ground state is given by
\[
U_{2}(\nu,x)=\sqrt{\frac{2\sin(\pi\varkappa)\,|E_{2}\left(  \nu\right)  |}%
{\pi\varkappa}}\,x^{1/2}K_{\varkappa}(|E_{2}\left(  \nu\right)  |^{1/2}x).
\]

According to \cite{GitTyV12}, the spectrum of $\hat{H}_{2,\nu}$ is given by
\textrm{Spec}$\hat{H}_{2,\nu}=\{E_{2}(\nu)\}\cup\lbrack0,\infty)$, and the
semiaxis $[0,\infty)$ is a continous part of the spectrum.

Setting $\mu=0,$ $s=s_{\min}(\nu)$ in (\ref{6.2.3.3}), we obtain the optimum
generalized oscillator representation for $\hat{H}_{2,\nu}$,$\,-\pi/2<\nu<0$,
bounded from below, but not nonnegative.

\subsubsection{$\alpha=-1/4$ ($\varkappa=0$), $\mu=\pi/2$}

A reasoning in this subsubsection is completely similar to those in subsubsec.
6.2.1 and subsubsec. 6.2.2.

According to (\ref{5.1.3.16}) and (\ref{5.1.3.18}), we have the identities
$\overline{\hat{b}}(\pi/2,s)=\hat{a}^{+}(\pi/2,s)$ and $\hat{b}^{+}(\pi/2,s)=$
$\overline{\hat{a}}(\pi/2,s)$, the point $s=0$ included, see subsec. 5.2, so
that with taking into account subsubsec. 6.1.3, formula (\ref{6.1.3.1}), we
find%
\[
\hat{H}_{3,\pm\pi/2}=\overline{\hat{b}}(\pi/2,s)\hat{b}^{+}(\pi/2,s)-(sk_{0}%
)^{2}\hat{I},\forall s\geq0,\,\varkappa=0,
\]
which is another form of the one-parameter family of oscillator
representations for $\hat{H}_{_{3,\pm\pi/2}}$ that is a restriction of the
known two-parameter-family of oscillator representations (\ref{6.1.3.1}) to
$\mu=\pi/2$. Of course, \ the comment following (\ref{6.1.3.1}) holds.

\subsubsection{$\alpha=-1/4$ ($\varkappa=0$), $0\leq\mu<\pi/2$}

A reasoning in this subsubsection is completely similar to that in subsubsec.
6.2.3 for the case of $0<\varkappa<1$.

We begin with representing the asymptotic behavior of functions $\phi
(\mu,s;x)$ (\ref{5.1.1.2}), $\mu\in\lbrack0,\pi/2)$, $s\in(0$,$\infty)$,
$\,\varkappa=0$, at the origin given by (\ref{5.1.1.4}) in a new form:%
\begin{align*}
&  \phi(\mu,s;x)=c[\sqrt{k_{0}x}\sin\theta(\mu,s)+\sqrt{k_{0}x}\ln(k_{0}x)
\cos\theta(\mu,s)]+O(x^{5/2}\ln x),\\
& \tan\theta(\mu,s)=\ln(s/2)-\tan\mu-\psi(1),\;
c=-\frac{\cos\mu}{\cos\theta(\mu,s)},\\
&  \mu\in\lbrack0,\pi/2),s\in(0,\infty),\,\varkappa=0,\,
\theta(\mu,s)\in\left(  -\frac{\pi}{2},\frac{\pi}{2}\right)  .
\end{align*}

We now determine the\ asymptotic behavior of functions belonging to the domain
$D_{H_{\mathfrak{e}b(\mu,s)}}$ of the operator $\hat{H}_{\mathfrak{e}b(\mu
,s)},$ $\mu\in\lbrack0,\pi/2),\,\,s\in(0,\infty),\,\,\varkappa=0,$ at the origin.

By definition of the operator $\hat{H}_{\mathfrak{e}b(\mu,s)}$, its domain
$D_{H_{\mathfrak{e}b(\mu,s)}}$ consists of functions $\chi\in D_{b^{+}(\mu
,s)}$ such that $\hat{b}^{+}(\mu,s)\chi=\eta\in D_{\overline{b}(\mu,s)}\subset
L^{2}(\mathbb{R}_{+})$. The first condition implies that $\chi$ allows
representation (\ref{5.1.3.8}) with $x_{0}=0$ and, in general, $D\neq0$, while
the second condition implies that $\eta(x)=O(x^{1/2}),\;x\rightarrow0$, see
(\ref{5.1.3.19}),\textrm{\ }(\ref{5.1.3.6}). Estimating the integral term in
(\ref{5.1.3.8}) with such $\eta$, we obtain that the asymptotic behavior of
functions $\chi\in$ $D_{H_{\mathfrak{e}b(\mu,s)}}$, $\mu\in\lbrack
0,\pi/2),\,s\in(0,\infty),\,\varkappa\in(0,1)$, at the origin is given by%
\begin{equation}
\chi(x)=C[(k_{0}x)^{1/2}\sin\theta(\mu,s)+(k_{0}x)^{1/2}\ln(k_{0}x)\cos
\theta(\mu,s)]+O(x^{3/2}),\,x\rightarrow0.\label{6.2.5.2}%
\end{equation}

According to \cite{GitTyV12}, for $\alpha=-1/4\,(\varkappa=0)$, there is a
one-parameter family of s.a. Calogero Hamiltonians with such asymptotic
behavior of functions belonging to their domains, namely , the family
$\{\hat{H}_{3,\nu},\,\nu\in\left(  -\frac{\pi}{2},\frac{\pi}{2}\right)  \}$.
The parameter $\nu$ is naturally identified with the angle $\theta(\mu,s)$ in
(\ref{6.2.5.2}), and we establish that%

\begin{align}
&  \hat{H}_{3,\nu}=\overline{\hat{b}}(\mu,s)\hat{b}^{+}(\mu,s)-(sk_{0}%
)^{2},\text{ }\nu=.\theta(\mu,s),\nonumber\\
&  \nu\in(-\pi/2,\pi/2),\,\mu\in\lbrack0,\pi/2),\,s\in(0,\infty),\,\varkappa
=0,\label{6.2.5.3}%
\end{align}
which represents a one-parameter family\textrm{\ }of generalized oscillator
representations for $\hat{H}_{3,\nu}$ with $\alpha=-1/4\,(\varkappa=0)$ and
$\nu\in(-\pi/2,\pi/2)$.

It is convenient to take $\mu$ as the independent parameter, then $s$ is
easily determined from the relation $\tan\vartheta(\mu,s)=\tan\nu$ to yield%
\begin{equation}
s=s(\mu,\nu)=2e^{\tan\nu+\tan\mu+\psi(1)}.\label{6.2.5.4}%
\end{equation}
For fixed $\nu$, the function $s(\mu,\nu)$ (\ref{6.2.5.4}) is monotonically
increasing from $s_{\min}(\nu)$ to $\infty$ as $\mu$ ranges from $0$ to
$\pi/2-0$, where%
\[
s_{\min}(\nu)=s(0,\nu)=2e^{\tan\nu+\psi(1)}.
\]

It is evident that the spectrum of $\hat{H}_{3,\nu}$ is bounded from below by
\[
-(s_{\min}(\nu)k_{0})^{2}=-4k_{0}{}^{2}e^{2(\tan\nu+\psi(1))}=E_{3}(\nu).
\]
Because $\ker\hat{b}^{+}(0,s_{\min}(\nu))=\{cK_{0}(s_{\min}(\nu)k_{0}%
x)\}\neq\{0\}$, whereas $\ker\hat{b}^{+}(\mu,s(\mu,\nu))=\{0\}$, $\,0<\mu
<\pi/2$, this boundary is an exact lower boundary of the spectrum of
$\hat{H}_{3,\nu}$, $E_{3}(\nu)$ is an eigenvalue of $\hat{H}_{3,\nu}$, the
energy of its negative ground level, and the normalized eigenfunction of the
ground state is given by%
\[
U_{3}(\nu,x)=\sqrt{2|E_{3}\left(  \nu\right)  |}\,x^{1/2}K_{0}(|E_{3}\left(
\nu\right)  |^{1/2}x).
\]
According to \cite{GitTyV12}, the spectrum of $\hat{H}_{3,\nu}$ is given by
\textrm{Spec}$\hat{H}_{3,\nu}=\{E_{3}(\nu)\}\cup\lbrack0,\infty)$, and the
semiaxis $[0,\infty)$ is a continuous part of the spectrum.

Setting $\mu=0$, $s=s_{\min}(\nu)$, in (\ref{6.2.5.3}), we obtain the optimum
generalized oscillator representation for $\hat{H}_{3,\nu}$, $\nu\in
(-\pi/2,\pi/2)$, bounded from below, but not nonnegative.

\vspace{1cm}
\begin{acknowledgement}
I.T. thanks RFBR Grand 11-01-00830 and B.V. thanks RFBR Grand 11-02-00685 for
partial support.
\end{acknowledgement}

\end{document}